\documentclass[12pt,preprint]{aastex}

\begin{document}

\title{V3885 Sagittarius: a Comparison with a Range of Standard Model Accretion Disks \footnotemark[1]}
\footnotetext[1]
{Based on observations made with the NASA/ESA Hubble Space Telescope, obtained at the
Space Telescope Science Institute, which is operated by the Association of Universities 
for Research in Astronomy, Inc. under NASA contract NAS5-26555, and the NASA-CNES-CSA
{\it Far Ultraviolet Explorer}, which is operated for NASA by the Johns Hopkins University
under NASA contract NAS5-32985} 

%% Use \author, \affil, and the \and command to format
%% author and affiliation information.
%% Note that \email has replaced the old \authoremail command
%% from AASTeX v4.0. You can use \email to mark an email address
%% anywhere in the paper, not just in the front matter.
%% As in the title, you can use \\ to force line breaks.

\author{Albert P. Linnell$^2$, Patrick Godon$^3$, Ivan Hubeny$^4$, Edward M. Sion$^5$,
Paula Szkody$^6$, and Paul E. Barrett$^7$}

\affil{$^2$Department of Astronomy, University of Washington, Box 351580, Seattle,
WA 98195-1580\\
$^3$Department of Astronomy and Astrophysics, Villanova University,
Villanova, PA 19085\\
$^4$Steward Observatory and Department of Astronomy,
University of Arizona, Tucson, AZ 85721\\
$^5$Department of Astronomy and Astrophysics, Villanova University,
Villanova, PA 19085\\
$^6$Department of Astronomy, University of Washington, Box 351580, Seattle,
WA 98195-1580\\
$^7$United States Naval Observatory, Washington, DC 20392\\
}

\email{$^2$linnell@astro.washington.edu\\
$^3$patrick.godon@villanova.edu\\
$^4$hubeny@as.arizona.edu\\
$^5$edward.sion@villanova.edu\\
$^6$szkody@astro.washington.edu\\
$^7$barrett.paul@usno.navy.mil\\
}

\begin{abstract}

A $\widetilde{\chi}^2$ analysis of standard model accretion disk synthetic spectrum 
fits to combined $FUSE$ and STIS spectra of V3885 Sagittarius, on an absolute flux basis,
selects a model that accurately represents the observed SED.
Calculation of the synthetic spectrum requires the following system parameters.
The cataclysmic variable secondary star 
period-mass relation calibrated by Knigge in 2007 sets
the secondary component mass. A mean white dwarf (WD) mass from the same study, that is consistent 
with an observationally-determined 
mass ratio, sets the adopted WD mass of $0.7M_{\odot}$, and
the WD radius follows from standard theoretical models.
The adopted inclination, $i=65{\arcdeg}$, is a literature consensus, and is subsequently supported by 
$\widetilde{\chi}^2$ analysis.
The mass transfer rate is the remaining parameter to set the accretion disk $T_{\rm eff}$ profile, and
the $Hipparcos$ parallax constrains that parameter 
to $\dot{M}=5.0{\pm}2.0{\times}10^{-9}~M_{\odot}~{\rm yr}^{-1}$ by a comparison with observed spectra.
The fit to the observed spectra adopts the contribution of a $57,000{\pm}5000$K WD.
The model thus provides realistic constraints on $\dot{M}$ and $T_{\rm eff}$ for a large $\dot{M}$
system above the period gap.
%The derived $\dot{M}$ represents that parameter value for a system above the period gap, 
%and the WD $T_{\rm eff}$ from the system model represents that parameter value for a large $\dot{M}$
%system.

\end{abstract}

%% Keywords should appear after the \end{abstract} command. The uncommented
%% example has been keyed in ApJ style. See the instructions to authors
%% for the journal to which you are submitting your paper to determine
%% what keyword punctuation is appropriate.


%% From the front matter, we move on to the body of the paper.
%% In the first two sections, notice the use of the natbib \citep
%% and \citet commands to identify citations.  The citations are
%% tied to the reference list via symbolic KEYs. The KEY corresponds
%% to the KEY in the \bibitem in the reference list below. We have
%% chosen the first three characters of the first author's name plus
%% the last two numeral of the year of publication as our KEY for
%% each reference.

\section{Introduction}

Cataclysmic variables (CVs) are semi-detached binary stars in which a late main sequence star 
loses mass onto a white dwarf (WD) by Roche lobe overflow \citep{w1995}. 
In non-magnetic systems the mass
transfer stream produces an accretion disk with mass transport inward and angular momentum 
transport outward, driven by viscous processes. The accretion disk may extend inward to the WD; the
outer boundary extends to a tidal cutoff limit imposed by the secondary star in the steady state case.
If the mass transfer rate is below a certain limit, the accretion disk is unstable
and undergoes brightness cycles (outbursts), and if above the limit, the 
accretion disk is stable against outbursts.
The latter objects (which have no recorded outburst of any type) are called nova-like (NL) systems.
As shown by current studies 
(\citet{Ribiero2007,Hartley2005,d2002,Woods1992,Metz1989,Haug1985,Guinan1982,Cowley1977}) V3885 Sgr is 
a NL system (see Warner 1995 for a detailed discussion).
V3885 Sgr has a $Hipparcos$ \citep{perry1997} parallax of $9.11 \pm 1.95$~mas. 
The orbital period is $0.20716071^{\rm d}$ \citep{Ribiero2007} and the inclination $i{\sim}65{\arcdeg}$
\citep{Haug1985,Hartley2005,Ribiero2007}.

NL systems are of special interest because they are expected to have an accretion disk radial temperature
profile	given by an analytic expression \citep[eq.5.41]{fr92} (hereafter FKR) which defines
the so-called standard model; the expression includes the mass transfer rate $\dot{M}$ as an
explicit variable. 
In NL systems, which are above the period gap \citep{howell2001}, the accretion disk
dominates the system spectrum. Fitting a synthetic spectrum, based on the analytic model, a proxy of
the accretion disk temperature profile, to an
observed spectrum potentially determines $\dot{M}$, a physical parameter of basic importance
since it controls the evolution of CV systems \citep{howell2001}. But the  analytic expression
also is an explicit function of the WD mass, $M_{\rm wd}$, and it must be determined independently.

By the basic paradigm of CV evolution \citep{howell2001} there is a close relation between
orbital period and donor mass. In a recent paper, \citet{knigge2007} has calibrated the
relations among $P$, $M_2$, $R_2$, $T_{\rm eff,2}$, and donor spectral type. Thus, given an
orbital perod below the tabular upper limit of $6^{\rm h}$, $M_2$ can be determined independently 
of other parameters. If there is an observationally-determined mass ratio,
the WD mass follows directly.

A normalizing factor is necessary in comparing synthetic spectra with observed spectra and the factor
depends on the distance to the system. If the distance is known independently an absolute determination
of $\dot{M}$ potentially is possible.

A complication in fitting a synthetic spectrum to an observed spectrum is the fact that the WD
may contribute detectably to the system spectrum; evaluating the WD contribution requires the WD $T_{\rm eff}$.
A limited number of CV systems have known values for their WD $T_{\rm eff}$ \citep{sion1999,winter2003}
with extremely few being NLs. 
Because of the large $\dot{M}$ in NL systems, the accretion disk dominates the system spectrum 
in the UV and optical, and makes 
determination
of the WD $T_{\rm eff}$ difficult.
Thus, the $T_{\rm eff}$ of an additional NL WD is of obvious interest.

\section{The $FUSE$ and STIS Spectra}

Table~1 presents the observations log. The STIS spectra are the same as
three STIS spectra described by \citet{d2002}.

\subsection{The $FUSE$ Spectrum}

$FUSE$ is a low-earth orbit satellite, launched in June 1999. Its optical
system consists of four optical telescopes (mirrors), each separately
connected to a different Rowland spectrograph. The four diffraction
gratings of the four Rowland spectrographs produce four independent
spectra on two microchannel plates (detectors). Two mirrors and two
gratings are coated with SiC to provide wavelength coverage below
1020\AA, while the other
two mirrors and gratings are coated with Al and LiF.  The
Al+LiF coating provides about twice the reflectivity of SiC at
wavelengths $>$1050\AA, and very little reflectivity below 1020\AA\ . 
These are known as the SiC1, SiC2, LiF1 and LiF2 channels
producing a total of 8 (partially overlapping) spectral segments:
SiC1a, SiC1b, SiC2a, SiC2b, LiF1a, LiF1b, LiF2a, LiF2b for each
exposure.

Four $FUSE$ observations of V3885 Sgr prospectively are available for analysis.
An observation on Aug. 10, 2002 has a short exposure time and has inferior S/N
ratio compared to other exposures. 
An exposure on 24 May 2000 (UT:01:51:55) has the longest exposure time and
occurs within 25 days of our chosen STIS spectrum. We chose this spectrum for analysis.
The exposure used 
the 30"x30" LWRS Large Square Aperture in TIME TAG mode.
The telescope collected 11 exposures obtained during 
11 consecutive orbits of the {\it FUSE} spacecraft, 
totaling 12865s of raw exposure time for the   
88 spectral segments; we downloaded the spectra from the MAST archive and processed
them as described below.  

The data were processed with the latest and final version of
CalFUSE (v3.2.1) \citep{dixon2007}.
The main change from previous versions of CalFUSE is that the data
are maintained as a photon list (the intermediate data file - IDF) throughout
the pipeline. Bad photons are flagged but not discarded, so the user can
examine, filter, and combine data without re-running the pipeline. 
Event bursts (short periods during an exposure when high count
rates are registered on one or more detectors) are processed 
automatically in this latest version of CalFUSE. 
The bursts exhibit
a complex pattern on the detector, their cause, however, still is unknown
(it has been confirmed \citep{dixon2007} that they are not detector effects).

The 11 exposures all have essentially identical continuum flux levels and there
is no detectable change in flux from exposure to exposure. The
main features that vary from exposure to exposure are the depth,
width and shape of the absorption lines.  
The next step is to coadd the 11 exposures, 
for the 8 individual segments.  
The count rate indicates screening of 22s of high voltage
(HV) for the 8 spectral segments.  
In addition there was a loss of good exposure time due to event
bursts of 472s, 409s, 390s, and 511s in segments 1a's, 1b's, 2a's and 2b's 
respectively. We obtain a total of 12310s, 12373s, 12392s and
12271s for segments 1a's, 1b's, 2a's and 2b's, respectively,    
including more than 8200s of NIGHT time (i.e. when the spacecraft is 
located in the shadow of the earth).   
Since the total {\it good collection time} varies, we   
weighted the segments accordingly (see below).

During
the observations, Fine Error Sensor A (which images the LiF1 aperture)
was used to guide the telescope. 
The spectral regions covered by the
spectral channels overlap, and these overlap regions are then used to
renormalize the spectra in the SiC1 (a \& b), LiF2 (a \& b), and SiC2  (a \& b)
channels to the flux in the LiF1 (a \& b) channel.
The low sensitivity portions (usually the edges)
of each channel are discarded. 
In addition, 
there exists a narrow dark stripe of decreased flux
in the spectra running in the dispersion direction,
{\it affectionally} identified as the ``worm'',
which can attenuate as much as
50\% of the incident light in the affected portions of the
spectrum, due to shadows of the wires in a grid
above the detector.
Because of the temporal changes in the strength and position of the
worm, CalFUSE \citep{dixon2007}
cannot correct target fluxes for its presence.
Therefore, we carried out a visual inspection of the {\it{FUSE}} channels to
locate the worm and we {\it{manually}} discarded the
portion of the spectrum affected by the worm. 
:$\sim$1130-1189\AA~in the LiF1b channel. 

We combined the individual exposures and channels to create a
time-averaged spectrum, weighting
the flux in each output datum by the exposure time (as mentioned earlier) 
and sensitivity of the
input exposure and channel of origin.
The final product is  a spectrum that covers almost the
full {\it{FUSE}} wavelength range 905\AA-1187\AA;
because we disregarded that segment of the spectrum affected by the
worm, the spectrum ends at $\sim 1182$\AA. 
The absorption lines in the final spectrum are blue shifted by about 0.4\AA.
Based on the \citet{Ribiero2007} ephemeris, the $FUSE$ spectrum	extends from orbital phase 
0.29 to orbital phase 1.01.

\subsection{The STIS spectra}

The STIS spectra (from the MAST archive) were taken in TIME-TAG mode, 
with the FUV-MAMA detector. 
The E140M echelle grating was used 
(with a central wavelength of 1425\AA), and the 0.2x0.2 aperture.  
The STIS spectra were processed with CALSTIS version 2.22. 
The calibrations were performed using the STSDAS package. 
Each STIS spectrum includes 44 echelle spectra, each one covering 
roughly 18\AA. The echelle spectra overlap except in the longer
wavelengths. As a consequence when the echelle spectra are merged to
produce the final spectrum, there are 4 gaps in the longer wavelengths.  
The gaps are about $\sim$0.5\AA\ wide and are at intervals of about
18\AA. We have removed the gaps by hand. 

The first and third STIS spectrum have closely similar continua  
and differ slightly in the shape of their absorption lines, with a 
blue shift of 1.3\AA~and 0.6\AA. The second
STIS spectrum has a continuum flux level 14\% larger and its absorption
lines  are blue shifted by only 0.4\AA.
Based on the ephemeris of \citet{Ribiero2007}, the 
Table~1 STIS spectra extend in orbital phase from 0.67 to 0.81, 0.33 to 0.47, 
and 0.35 to 0.49 respectively. The STIS spectra cover the interval from 1150\AA~
to 1720\AA.  

\subsection{The combined spectra} 

The $FUSE$ spectrum and the first STIS spectrum of V3885 Sgr are shown in Figure~1
before correction for reddening; 
the spectra are characterized by broad and
deep absorption lines from the source (disk plus wind/chromosphere). These broad absorption lines
are blue shifted by about 1.5\AA\ and 2.5\AA\ in the $FUSE$ and  
STIS spectra respectively; they are about 4\AA\ 
in width ($\sim$1,000km/s) 
and exhibit the same asymmetry in both the $FUSE$ and STIS spectra. 
The red edge of the absorption is steep, while the blue edge has a more 
gradual slope, resembling the blue-shifted absorption part of a 
P-Cygni profile. However, only the C\,{\sc iv} (1550) and N\,{\sc v} (1240) 
lines   
exhibit a true P-Cygni profile, while all the other broad absorption lines do not.  
All these lines belong to high ionization species: 
N\,{\sc iii}, N\,{\sc iv},  N\,{\sc v},  
Si\,{\sc iii}, Si\,{\sc iv}, 
C\,{\sc iii}, C\,{\sc iv}, 
O\,{\sc vi}, P\,{\sc v}, and  He\,{\sc ii}. 

In addition to the broad lines, there are many narrow and deep absorption lines.
They have a 
width varying between about 0.15\AA\ (e.g. O\,{\sc i}) and 0.3\AA\ 
(for H\,{\sc i}) and are also blue shifted, but by a much smaller
amount ($\sim 0.2$\AA\ ) than the broad lines. Many of the narrow lines belong to low order ionization
species; however, some also belong to higher ionization species. 
Some of the sharp lines are known to
be interstellar, such as Ar\,{\sc i} and Al\,{\sc ii}.
For simplicity we put all these lines in the ISM category.
Additional details concerning the STIS spectra are in \citet{d2002}. 
Table~2 and Table~3 list all the lines, separated into
$FUSE$ lines and STIS lines.  

From the AAVSO web site\footnote{www.aavso.org} we determine that 
during the month of April 2000 and through May 2000, the V3885 Sgr
visible magnitude varied by only 0.2. The first of the Table~1 STIS spectra falls
within this time frame and it is closest in time to the $FUSE$ spectrum.
Also, the orbital phases of the two spectra are nearly the same. Finally, the
first STIS spectrum most closely matches the $FUSE$ spectrum in the overlap region.
All of the spectra have been corrected for interstellar reddening of E(B-V)=0.02 (\S3).
We choose the first STIS spectrum to combine with the $FUSE$ spectrum for analysis. 
Figure~2 shows the combination of the $FUSE$ spectrum with the three STIS spectra.
The upper panel shows the overlap region. The $FUSE$ spectrum is red, the first STIS
spectrum is cyan, the second STIS spectrum is magenta, and the third STIS spectrum
is brown. For optimum fit to the $FUSE$ spectrum, the STIS spectra have been divided
by 0.98, 1.12, and 0.94 respectively. The top panel shows the very good match of the
spectra in the overlap region, while the lower panel shows that, when normalized,
the three STIS spectra track each other with excellent fidelity. 
The STIS spectra are noisy in the overlap region while the upper panel of Figure~2
shows that the STIS and $FUSE$ spectra continua superpose accurately. We produced a
combined spectrum for analysis by terminating the STIS spectrum at 1181.5\AA~at
the short wavelength end, 
dividing it by 0.98, and adding the $FUSE$ spectrum, beginning at 1181.5\AA~and extending
to shorter wavelengths.

We suggest, without
detailed analysis, that the differences among the STIS spectra could be explained by
slight variation in the mass transfer rate from the secondary star. Whatever the cause
of the variation, the fact that the normalized STIS spectra have the same SED simplifies
calculation of a system model.

\section{ Initially adopted system parameters}

Table~4 (and notes) lists initially adopted parameters and their sources. Several of the system parameters are 
poorly known.
As mentioned earlier, the Hipparcos parallax is $9.11{\pm}1.95~{\rm mas}$, which corresponds to
a distance of $110{\pm}30~{\rm pc}$.
\citet{knigge2007} determines $M_2=0.475M_{\odot}$ for $P=0.20716^{\rm d}$, and a mean 
$M_{\rm wd}=0.7M_{\odot}$. We adopt both masses and note that the resulting $q=0.68$ is in
agreement with the observationally determined $q{\sim}0.7$ \citep{d2002}.
\citet{panei2000} lists a WD radius of $1.091{\times}10^{-2}R_{\odot}$
for a $0.70M_{\odot}$ homogeneous zero temperature Hamada-Salpeter carbon model WD.
In our subsequent study of the observed spectra we correct the radius for the adopted WD $T_{\rm eff}$;
the contribution of the WD to the system synthetic spectrum is appreciable.
Tablenotes to Table~4 list alternative sources for the adopted initial parameters.
Those notes do not include the determination of $i{\sim}25{\arcdeg}-30{\arcdeg}$ \citep{Cowley1977}.
This determination was based on radial velocity observations with a large scatter; the derived
velocity amplitude was $K=74~{\rm km}~{\rm s}^{-1}$ while much more accurate data gives a value
of $K=165~{\rm km}~{\rm s}^{-1}$ \citep{Hartley2005}. 

\citet{ver1987} 
determined a
rough upper limit of $E(B-V)$ of 0.04 (and a preferred value of $E(B-V)=0.0$) for V3885 Sgr based on a study 
of the 2200\AA~``bump" in $IUE$ spectra. 
\citet{bruch1994} list $E(B-V)=0.02$, which we adopt.

A number of studies have considered the tidal cutoff boundary, $r_d$, of accretion disks
(\citet{pac1977}; \citet{pp1977}; \citet{wh1988}; \citet{scr1988}; \citet{wk1991}; \citet{g1993}).
These authors agree on $r_d{\approx}0.33D$, where $D$ is the separation of the stellar components.
We adopt this expression for the tidal cutoff radius of the accretion disk.

\section{The analysis program: BINSYN}

Our analysis uses the program suite BINSYN \citep{linnell1996}; recent papers \citep{linnell2007,linnell2008a}
describe its
application to CV systems in detail. 
Briefly, an initial calculation produces a set of annulus
models for a given WD mass, radius, and mass transfer rate. This calculation uses the program
TLUSTY~\footnote{http://nova.astro.umd.edu} \citep{hubeny1988,hl1995}. 
The program considers the radial and vertical structure of the disk independently;
the radial structure is based on the standard model (FKR) and so follows the prescribed relation
between local $T_{\rm eff}$ and the annulus radius. The vertical structure is solved,
self-consistently, as described by \citet{hubeny1990} and \citet{hh1998}.
The set of annulus models covers the accretion disk from its innermost (WD) radius to 
just short of its outermost 
(tidal cutoff) radius.

The primary source of viscosity in CV accretion disks
is MRI \citep{bh1991}. \citet{hirose2006} calculate MHD models with local dissipation of turbulence and
show that the vertical extent of an annulus is greater than in previous models.
\citet{blaes2006} show that magnetic support has a significant effect on synthetic spectra of
black hole (BH)
accretion disk annuli and illustrate the effect in the case of a BH system with 
$M_{\rm BH}=6.62M_{\odot}$ and with an adopted $\alpha=0.02$.	
They are able to simulate the effect of magnetic support within the TLUSTY framework
by adjusting the TLUSTY parameters $\zeta_0$ and $\zeta_1$ and use this simulation to calculate
the effect on synthetic spectra. The primary effect is a hardening of the spectrum.
\citet{king2007} point out that MHD models that produce viscosity via MRI require $\alpha$ values
that are a factor 10 smaller than the $\alpha$=0.1-0.4 required by observational evidence and
suggest that some caution still is needed in accepting the MHD results.	Our models have used the
default values $\zeta_0=\zeta_1=0.0$ and we assume the effect of magnetic support on our
annulus spectra, with $M_{\rm wd}=0.7M_{\odot}$, is insignificant.

Table~5 lists properties of annuli calculated with TLUSTY(v.202)
for a mass transfer rate of $5.0{\times}10^{-9}M_{\odot}/{\rm yr}^{-1}$.
This value initially was chosen for test purposes because of the similarity of V3885 Sgr to
IX Vel \citep{linnell2007}; see \S5 for a detailed discussion. 
The control file to calculate a given annulus requires a radius of the WD in units of $R_{\odot}$.
All of the annuli used the radius of a zero temperature WD for a homogeneous 
carbon Hamada-Salpeter
$0.70M_{\odot}$ model from \citet{panei2000}.
All of the annuli are solar composition H-He
models, and the models through $r/r_{\rm wd,0}=20.00$ are converged LTE models. The remaining annuli are
LTE-grey models (see the TLUSTY Users Guide for an explanation).
A \citet{ss1973} viscosity parameter $\alpha=0.1$ was used in calculating all annuli.
Each line in the table 
represents a separate annulus. Temperatures are in Kelvins.
The column headed by $m_0$ is the column mass, in ${\rm gm}~{\rm cm}^{-2}$, 
above the central plane.
The columns headed by $z_0$ and $N_e$ are, respectively, the
height (cm) above the central plane for a
Rosseland optical depth of ${\approx}0.7$ and the electron density (${\rm cm}^{-3}$) at the same level. 
The column headed
by ${\rm log}~g$ is the log gravity (cgs units) in the $z$ direction at a Rosseland optical depth of ${\approx}0.7$. 
The column headed by ${\tau}_{\rm Ross}$
is the Rosseland optical depth at the central plane. We call attention to the fact that the annuli are optically
thick to the outer radius of the accretion disk.

Following calculation of annulus models for assigned $M_{\rm wd}$ and $\dot{M}$,
program SYNSPEC(v.48)~\footnote{http://nova.astro.umd.edu} \citep{hlj1994} is used to produce a 
synthetic spectrum for each annulus.
We adopted solar composition for all synthetic spectra. The synthetic spectra include contributions
from the first 30 atomic species, assumed to be in LTE.

We set SYNSPEC to produce light intensities at 10 zenith angles (equally spaced in $cos(\gamma)$ between
1.0 and 0.1,
where $\gamma$ is
the angle between the given direction and the local normal) at each calculated wavelength for
all the individual annulus synthetic spectra. This choice automatically accounts for wavelength-dependent
limb darkening in calculating synthetic spectra at different orbital inclinations.
BINSYN calculates a synthetic spectrum for each of the annuli specified in the BINSYN system model by
interpolation among the array of TLUSTY annulus models, with proper allowance for inclination.

BINSYN models the complete CV system, including the WD, secondary star, accretion disk face,
and accretion disk rim as separate entities. The model represents phase-dependent and 
inclination-dependent effects, including eclipse effects and irradiation effects, on all of the system objects. The
representation of the stars requires polar $T_{\rm eff}$ values and gravity-darkening exponents.
BINSYN represents the accretion disk by a specified
number of annuli, where that number typically is larger than the number of TLUSTY annulus
models. The accretion disk $T_{\rm eff}$ profile may follow the standard model, but, alternatively, the profile
may be specified by a separate input file. 

BINSYN has provision to represent a bright spot on the rim face, 
but the low amplitude $V$ light curve by \citet{Ribiero2007}(Fig.~10) is consistent with light variation
of an irradiated secondary star so 
we do not include a rim bright spot. Based on the analysis by \citet{smak2002} we set the rim $T_{\rm eff}$
equal to the $T_{\rm eff}$ of the outermost annulus. 

\section{A V3885 Sgr system model}

We adopted system parameters listed in Table~4 for an initial test.
Based on the similarity to IX Vel (comparable periods, visual magnitudes, parallaxes, CV class,
spectroscopic properties \citep{d2002}), we first tried a standard model accretion disk with
the same $\dot{M}$, $5.0{\times}10^{-9}~M_{\odot}~{\rm yr}^{-1}$,
as for IX Vel \citep{linnell2007}, and, as for IX Vel,
the (initial) model included a 60,000K WD. In constructing a model accretion disk
it is necessary to allow for the change in WD radius from the
zero temperature model. We used Table~4a of \citet{panei2000} to estimate the 60,000K WD radius at 
$0.0134R_{\odot}$. Our BINSYN model specified 33 annuli for the accretion disk, with assigned
radii and corresponding standard model $T_{\rm eff}$ values listed in Table~6. The larger hot WD
radius than a zero temperature WD radius at a given multiple of the `WD radius'
produces a lower Table~6 temperature 
than Table~5 (e.g., at the $r/r_{\rm wd}$ value 6.0 in column~1 of both tables). The synthetic spectrum
for a given BINSYN annulus is calculated by interpolation (temperature-wise) among the Table~5 entries. 
Ideally, we could recalculate the whole series of annulus models (via TLUSTY) based on the new value of 
$R_{\rm wd}$,
but the change in the individual annulus models would be very small and would not be warranted in view
of other uncertainties (e.g., the system parallax).

Table~5 ends
at a $T_{\rm eff}$ of 7711K. TLUSTY fails for larger radii, even in the gray model case. The net
effect is that the calculated BINSYN outer annuli have synthetic spectra corresponding to 7711K even
if the specified standard model calls for a lower temperature. The effect on the synthetic spectrum 
of failing to follow the standard model is negligible. Adding
a lower temperature (3500K) stellar synthetic spectrum to the array of Table~5, thereby permitting
interpolation to a lower temperature in the outer BINSYN annuli, made no detectable change
in the system synthetic spectrum. 
As separate issues, the outer annuli temperatures should not fall below 6000K
to avoid outbursts \citep{osaki1996,lasota2001} and tidal dissipation is expected to raise the outer
accretion disk temperature above the standard model value \citep{smak2002}.

The overall SED fit to  the combined spectrum of \S2.3, using initial parameters, and judged by visual estimate, 
was very good with a standard model accretion 
disk temperature profile.
This was in contrast to IX Vel which required modification of the temperature profile from a
standard model to achieve
an acceptable fit \citep{linnell2007}.
Assessing the quality of fit for V3885 Sgr is complicated by the presence of strong absorption 
features of highly
ionized material (Figure~1), probably associated with a wind \citep{d2002}. 
At this initial stage, the extreme synthetic spectrum FUV region had slightly too large flux: 
the only parameter (either $\dot{M}$ or $T_{\rm eff,wd}$) available to
adjust the flux while preserving the fixed normalizing factor was the WD $T_{\rm eff}$. 
($\dot{M}$ could be adjusted but that route produced detectably larger residuals longward of 1500\AA.
Small adjustments of $\dot{M}$ move the entire synthetic spectrum up or down without appreciable change
in the spectral gradient. Similarly, a change in $i$ does not change the spectral gradient.
We consider a boundary layer in \S6.)
We reduced the
WD $T_{\rm eff}$ to 57,000K and found this produced an optimum fit as judged by eye estimate.
The new model neglected the very small change in WD radius
between $T_{\rm eff}=60,000$K and 57,000K, which would have affected the tabular radii in Table~6.

Figure~3 shows the FUV fit to the combined $FUSE$ and STIS spectra, dereddened for ${E(B-V)=0.02}$.
We call attention to the broad absorption lines which we attribute to a wind/chromosphere and that 
affect a major part of
the $FUSE$ spectrum. Note the deep ISM Lyman series lines, the broad and deep absorption by OI and NIII
near 990\AA, the deep and possible multi-component Ly$\beta$ line near 1025\AA, and the broad lines of SiIII, PV, 
SIV and SiIV in the 1110\AA~to 1140\AA~interval. The overlapping and possibly multicomponent lines
depress the apparent continuum over the entire interval from 970\AA~to 1170\AA.
   
The normalizing factor to divide the synthetic spectrum to superpose it on the observed, dereddened
combined $FUSE$ and STIS spectrum, corresponding to an exact distance of 110 pc, is 
$1.1473{\times}10^{41}~{\rm cm}^2$. We stress that this is not an adjustable parameter. In \S5.1 we consider
the effects of varying the `exact' distance corresponding to ${\pm}1{\sigma}$ in the $Hipparcos$ parallax.
The
spectrum at the bottom is the contribution of a 57,000K WD. For simplicity we assume the WD is rotating
synchronously; we do not fit any observed absorption lines in detail.
The program assumes that the accretion
disk blocks the view of the lower half of the WD. In addition, BINSYN allows for the partial
visibility of that part of the WD not obscured by the accretion disk and above the observer's horizon, 
and includes wavelength-dependent
limb darkening. The spectral resolution of all the synthetic spectra is 0.02\AA~(\S5.2.).
The adopted orbital phase for the synthetic
spectra was 0.75, in agreement with the observed spectra, \S2. Adopting an orbital phase of 0.25
produced a clear misalignment of the calculated 960\AA~peak with the observed one.

We check the apparent best fit, by eye estimate, with a reduced chi squared
($\widetilde{\chi}^2$) test. 
We did not smooth the observed spectrum for this test.
The presence of unmodeled
wind/chromosphere features in the observations produces large residuals (especially in the FUV, Figure~3).
A $\widetilde{\chi}^2$ test of our model fit to the observed continuum requires masking those spectral regions excluded 
from the fit (i.e., the wind/chromosphere features). 
We initially masked spectral lines by hand, including the deep Lyman lines. (The Lyman lines in the model,
mostly from the accretion disk, are washed out by Keplerian rotation.) This first stage masking produced a
major reduction (from 154.39 to 7.38) in $\widetilde{\chi}^2$ as compared with the unmasked spectrum. 
There remained a major "valley"
in a residuals plot, described in the following paragraph, with the largest (negative) residuals near 1100\AA.
Our previous discussion has identified this effect with overlapping absorption features from a
wind/chromosphere which depresses the entire spectral interval over the approximate range 970\AA~to 1170\AA.
BINSYN has no present capability to model a wind/chromosphere and further progress requires masking
the affected spectral interval.
We therefore masked the entire region from
970\AA~to 1170\AA, producing a further large reduction in $\widetilde{\chi}^2$ (from 7.38 to 2.65). 

Figure~4 shows plots 
of the residuals in the three cases. 
The top plot, displaced
upward by 12.0 ordinate units, is the residuals for the unmasked case. The middle plot, displaced upward
by 4.0 ordinate units, is the partially masked case. 
The lower (undisplaced) plot is the final set of residuals adopted for this study. 
In each case a horizontal line marks the undisplaced
reference.
The change from the top to middle plot mostly results from masking of individual spectral features,
such as the Lyman lines, the CIV $\lambda$1548 feature, etc. 
The STIS spectrum appears to have a defect beginning at 1710\AA, and the three STIS spectra show
more variability longward of the HeII $\lambda$1640 line than in other spectral regions.
We terminated our calculated fit just longward of the HeII $\lambda$1640 line.	
The middle plot clearly shows the residual depression after individual spectral
features have been masked and the bottom plot shows the overall improvement in residuals after
masking the 970\AA~to 1170\AA~region. The improvement is even more apparent on individual plots
with extended ordinate axes.

The lower plot still has fairly large outliers.
Although further selective masking could be done,
we consider it dangerous to go very far in suppressing observations that disagree with the model.
Residual absorption features could be masked but there are positive outliers that then would bias the solution.
We
anticipate that it will be possible to locate an optimum model with the chosen masked set of observations. 
The final
masked version of the observed combined spectra includes 4332 tabulated wavelengths and corresponding
flux values.

It is particularly unfortunate for this analysis that most of the $FUSE$ spectrum must be masked,
but it is unavoidable: the masked region clearly is strongly affected by the broad and deep absorption
lines of high excitation species as described in \S2.3. The high excitation involved excludes the
lines from production by the accretion disk itself or the WD. Further, ignoring the high excitation
aspect for the moment, it would not be possible to simulate the depressed 970\AA-1170\AA~spectral interval
by adopting a modified accretion disk $T_{\rm eff}$ profile. Our experience with non-standard model
accretion disks is that the calculated SED profile can have its slope changed over an extended
spectral region but it is not possible to put a "hole" in the SED.  

We calculated $\widetilde{\chi}^2$ values for a grid of 34 models in the $\dot{M}$, $T_{\rm eff,wd}$ plane, 
centered on
the best visual estimate model fit to the masked observations.
Table~7 lists those values plus $\widetilde{\chi}^2$ values for 4 additional $i$ values (to be discussed separately).
The aim is to provide a sufficiently extended grid that a subset of $\widetilde{\chi}^2$ 
values define a  $\widetilde{\chi}^2$ contour that encloses
the region of minimum $\widetilde{\chi}^2$, given the adopted values of other parameters.
Calculation of a given model is a time-consuming process and a compromise is necessary between total
computational time and density of grid points. Our computed grid is sparse but our judgment is that it
meets our needs, given the uncertainties of separately adopted parameters.	
%Figure~5 shows the $\widetilde{\chi}^2$ values on the $\dot{M}$, $T_{\rm eff,wd}$ plane.
%Note that a $\widetilde{\chi}^2$ value of 4.0 or greater bounds the tabulated values in any horizontal,
%vertical or diagonal trace through the tabulated region. This fact is the basis for our statement that
%the grid density is sufficient for our purposes.
We have used the IDL routine CONTOUR to produce contour plots for the Table~7 $i=65{\degr}$ values.
The $\dot{M}=5.0{\times}10^{-9}~M_{\odot}~{\rm {yr^{-1}}}$,
$T_{\rm eff,wd}=57,000$K model has
the smallest tabular $\widetilde{\chi}^2$ value and we take that model to be our solution for V3885 Sgr.
We discuss this choice and the topic of grid density, including its adequacy, in more detail in \S6.	

We have sampled the $\dot{M}$, $T_{\rm eff,wd}$ plane for the case $M_{\rm wd}=0.8M_{\odot}$ with two new models:  
$\dot{M}=5.0{\times}10^{-9}~M_{\odot}~{\rm {yr^{-1}}}$ and $T_{\rm eff,wd}$=57,000K, producing
$\widetilde{\chi}^2$=3.06, and $\dot{M}=5.0{\times}10^{-9}~M_{\odot}~{\rm {yr^{-1}}}$ and
$T_{\rm eff,wd}$=52,000K, producing $\widetilde{\chi}^2$=3.81. The corresponding $\widetilde{\chi}^2$
values from Table~7 are 2.65 and 3.39 respectively. A preferred value of $M_{\rm wd}$
is not possible by $\widetilde{\chi}^2$ analysis of the existing data. The SED accretion disk model is 
not a strong function
of $M_{\rm wd}$ for the change from 0.7$M_{\odot}$ to 0.8$M_{\odot}$. Said differently, Figure~5 remains
roughly constant, with a small systematic shift for changes in $M_{\rm wd}$ of about 0.1 or less.

An improved value of $M_{\rm wd}$ may be possible with further optical region spectroscopy.
\citet{Ribiero2007} find that the H$\alpha$ emission line arises from irradiaton of the secondary while
the (doubled) $\lambda6678$ HeI emission line associates with the accretion disk.
If an accretion disk model that produces secondary component emission lines were available, a fit to the HeI line could
constrain $i$.

Table~7 includes 4 models for different assumed values of $i$ and with other parameters the same as the preferred model. 
There is a
very sharp minimum of $\widetilde{\chi}^2$ at the adopted $i=65\arcdeg$. At this inclination there is an
eclipse	of the outer accretion disk at orbital phase 0.0 with photometric effects much smaller than the predicted
variation due to irradiation of the secondary. Increasing the inclination to $70\arcdeg$ 
greatly deepens the depth of accretion disk eclipse and barely
fails to eclipse the WD; it would produce light curve effects that are not observed \citep[Figure~10]{Ribiero2007}.
However the observed light variation is complex and may include a contribution from a rim bright spot. 
An inclination of $71\arcdeg$ would eclipse the WD and produce a major light curve variation.
It is not possible to use the $\widetilde{\chi}^2$ $i$-dependent variation to determine $i$. 
The calculated system flux is
a strong function of $i$ because of the changing projection of the flux-dominating accretion disk on the
plane of the sky. The fit of the model spectrum to the observations requires a normalizing factor that is
given by the $Hipparcos$ parallax. The sharp minimum of $\widetilde{\chi}^2$ for $i= 65\arcdeg$ results from
the joint consistency of that parallax and the adopted $i$ value. Actual determination of $i$ would require
observational effects intrinsic to the system, i.e., eclipses.

Until a model that includes simulation of the wind/chromosphere is available it is doubtful that
substantial further reduction in $\widetilde{\chi}^2$ can be achieved.
Based only on the $\widetilde{\chi}^2$ plot, Figure~5, we estimate the
parameter errors in $\dot{M}$, $T_{\rm eff,wd}$, and $i$ as $2.0{\times}10^{-10}~M_{\odot}~{\rm {yr^{-1}}}$, 
$3000{\rm K}$, and, from Table~7, $2.0\arcdeg$ respectively. However, shifts in $\dot{M}$ to match the
uncertainty in the $Hipparcos$ parallax (\S5.1.) require a larger $\dot{M}$ error of
$2.0{\times}10^{-9}~M_{\odot}~{\rm {yr^{-1}}}$ which we adopt for the system solution. 
Further, in recognition of the sparse grid of models
and their associated $\widetilde{\chi}^2$ values, we increase the estimated $T_{\rm eff,wd}$ error to 5000K.

Our final system parameters are in Table~8.	Where appropriate we have provided estimated error limits.
Figure~6 shows the final model compared with the combined observations (not the masked observations).
The blue line is the contribution of the 57,000K WD. The lower of the two black lines is the contribution
of the accretion disk, and the upper black line is the system model. 

\subsection{The effects of $Hipparcos$ parallax uncertainty}

It is possible to test the effect of uncertainty in the $Hipparcos$ parallax by changing the
normalizing divisor for superposing the synthetic spectrum on the combined $FUSE$ and
STIS spectra. Changing the parallax by $1\sigma$ to 7.16~mas changes the normalizing divisor
from $1.147{\times}10^{41}~{\rm cm}^2$
to $1.857{\times}10^{41}~{\rm cm}^2$. The synthetic spectrum is displaced downward by approximately
two flux units on Figure~3 (the 1000\AA~peak moves from an ordinate value of 10. to 8.) 
Compensating for the change would require increasing $\dot{M}$
to perhaps $6.5$ or $7.0{\times}10^{-9}~M_{\odot}~{\rm yr}^{-1}$.
Correspondingly, changing the parallax to 11.06~mas changes the normalizing divisor to
$7.784{\times}10^{40}~{\rm cm}^2$ and displaces the synthetic spectrum upward by about two flux units,
from a 1000\AA~peak of 10 to 12.
Compensating would require decreasing $\dot{M}$ 
to perhaps $4.0$ or $4.5{\times}10^{-9}~M_{\odot}~{\rm yr}^{-1}$.

\subsection{Correction for the ISM}

Recent papers have included corrections for line absorption by the ISM \citep{godon2008,linnell2008a}.
A custom spectral fitting package is used to estimate the temperature and density of the
interstellar absorption lines of atomic and molecular hydrogen. The ISM model assumes that the temperature,
bulk velocity, and turbulent velocity of the medium are the same for all atomic and molecular species,
whereas the densities of atomic and molecular hydrogen, and the ratios of deuterium to hydrogen and
metals (including helium) to hydrogen are adjustable parameters. The model uses atomic data of
\citet{morton2000, morton2003} and molecular data of \citet{abgrall2000}. The molecular hydrogen 
transmission values have been checked against those of \citet{mccandliss2003}.
The calculation requires fairly high spectral resolution and we have calculated all of the
TLUSTY-SYNSPEC synthetic spectra at a resolution of 0.02\AA; we used the same resolution in
calculating the system synthetic spectra. 
The ISM parameters are in Table~9.
Figure~7 shows the corrected version of part of Figure~6. The ISM model produces a good representation of
the observed deep cores of the Lyman series lines and the Lyman cutoff, as well as some lines in other 
spectral regions
and clearly shows that the ISM produces the observed Lyman lines.
The broad, high excitation  absorption
regions from the accretion disk corona,
chromosphere, or wind, identified in Table~2 and Table~3, remain as prominent residuals.
As stated above, the Lyman lines were masked in the fit to the model and the spectral region shortward
of the Lyman limit was masked so only the Lyman region continuum participated in the model solution.
This was proper since a fit to the ISM does not affect the choice of the model parameters listed in Table~8.
There are small positive residuals from the model near the short wavelength end. This could be interpreted as
evidence for an unmodeled very high temperature contribution, discussed in the following section.

\section{Discussion}

In a well-known paper \citet{wade1988} showed that neither steady state model accretion disks based
on Planck functions or stellar model atmospheres could simultaneously fit the colors and absolute
luminosities of a set of NL systems (V3885 Sgr was included in the study). Tomographic analyses of 
accretion disks \citep{rut1992}
show clear departures from standard model temperature profiles (FKR).
Recent analyses of the NL systems MV Lyr, IX Vel, QU Car, and UX UMa 
\citep{linnell2005,linnell2007,linnell2008a,linnell2008b} 
showed, in each case,
that a standard model accretion disk synthetic spectrum could not accurately fit observed spectra.
These results suggest that physical conditions in possibly all NL systems are so complex, with winds,
coronae, iron curtains, etc., that an accurate fit by a standard model synthetic spectrum is too
much to expect. Consequently,
the most interesting result of this study is that a standard model synthetic spectrum plus WD accurately
fits the observations ($\widetilde{\chi}^2$=2.65) after masking the effects of a wind/chromosphere. 
This conclusion is particularly true in light of the fact that the system distance is known so
the fit is in absolute flux units. 
It is worth emphasizing that the model presented here includes a measurable contribution from a hot
source, in addition to the accretion disk, sufficiently
large that limits can be set on the hot source $T_{\rm eff}$. Assuming that the hot source
is a WD (see below),
this is the first time the effective temperature of a hot WD has been constrained during the high 
brightness state of a nova-like variable.

As discussed by FKR, half of the potential energy liberated in the fall of the mass transfer stream
from the L1 point (essentially from infinity) to the WD surface appears as radiated energy. Energy 
conservation requires an accounting for the other half and the presence of a hot boundary layer (BL) is
a common prescription. The problem is that BLs with the prescribed properties are not
observed \citep{cordova1995}. 
The absence of an observed but predicted (FKR) BL in high $\dot{M}$ CV systems has an 
extensive history: 
\citep{ferland1982,kaljen1985,
patray1985,hdr1991,hdr1993,vrt1994,idan1996}. 
One proposed explanation is that the WD is rotating close to Keplerian rotation, with no
BL predicted. However, HST measurements
starting with \citet{Sion1994} have measured rotational WD velocities too low to explain "missing"
BLs.
An alternative explanatory theme is that the absence of an
observed BL
associates with a wind, and in V3885 Sgr a wind is
clearly present \citep{d2002,Hartley2005}.

BINSYN has an option to simulate a BL by specifying the innermost annulus radial thickness and $T_{\rm eff}$. 
We applied the theory of \citet{godon1995} to obtain those two parameters. This
theory derives a BL with a $T_{\rm eff}{\sim}100,000$K and radial thickness of about $0.1R_{\rm wd}$.
This BL $T_{\rm eff}$ is consistent with a value from wind ionization \citep{hdr1991} and the theory
of \citet{idan1996}.
The \citet{godon1995} model adopted $M_{\rm wd}=1.0M_{\odot}$ and $T_{\rm wd,eff}$=20,000K.
Addition of the BL made a nearly negligible change in our system SED. There was a barely detectable
increase in the system flux at the short wavelength limit over the SED of our original model, producing
a slightly better fit in that spectral region. The evidence in favor of a BL is inconclusive; a model
including a 100,000K BL and a 57,000K WD is consistent with the observations.
\citet{popham1995} calculate BL models and illustrate their theory for the case
$\dot{M}=10^{-8}~M_{\odot}~{\rm yr}^{-1}$ and zero WD rotation, leading to a somewhat higher BL 
$T_{\rm eff}$. 
In a discussion among the coauthors Godon noted that the BL $T_{\rm eff}$ might be as high as 200,000K but 
probably would not be visible at $i=65{\arcdeg}$
because of an inflated inner disk; an additional reason for not including a BL.
A model including the 100,000K BL but with our WD contribution suppressed (i.e., requiring the hot source to
be the BL exclusively)
would be strongly inconsistent with
the observations.
We are unaware of any BL models
that treat MHD effects.

Is it possible to exclude a different source for the WD contribution of our present model?
We seek some spectral feature that could distinguish the WD contribution from a different prospective
source. 
In Figure~6 the WD Ly $\gamma$ absorption line appears to make an important contribution to the system
synthetic spectrum. A different smoothly varying source, such as a black body (BB) or
power law source might produce a
much poorer fit to the observed Ly $\gamma$ wings and so be excludable. We tested this possibility by 
replacing the WD synthetic
spectrum by a 57,000K BB with the same size as the replaced WD. The existing model uses
light intensities at 10 zenith angles (\S4), produced by SYNSPEC, for each annulus. Use of the BB spectrum
required a change to flux-based synthetic spectra with an adopted wavelength-independent limb darkening
coefficient; we chose 0.6. The system synthetic spectrum SED was essentially unchanged but required a slight
adjustment of the normalizing factor to reproduce the same fit as  with the original model, likely a result
of the change to a wavelength-independent limb darkening coefficient. The fit
at the short wavelength limit was slightly better, but it is uncertain whether this arises from the
BB replacement or the change in limb
darkening. In any event, the Ly $\gamma$ line was slightly shallower in the new simulation, arguing 
weakly in favor of the
presence of a WD, but the fit to the
wings was essentially the same. The proposed spectral feature test fails and we are unable to exclude a
BB as the source of the required hot addition to the accretion disk.
The SED of
a power law source would differ from that of a BB source and should produce a poorer fit to the
observed spectrum; without explicit test we discount this type source as a reasonable candidate.

To answer the question ``What is the source of the hot addition to the accretion disk?" the most reasonable
working hypothesis is that the hot source is a WD.
A WD is known to be present (it is a CV system)
and the inclination, $i$, is such that the WD should be visible. 
A spectral source substitute for a WD (e.g., BB) would need not only to explain why the WD spectrum is not seen but
also why a BB source of the same $T_{\rm eff}$ and size (which sets the radii of the accretion disk annuli
and thus their $T_{\rm eff}$ values (Table~6))
and physical location (exactly centered on the accretion disk) is a reasonable substitute.

The adequacy of our set of 34 models in populating the $\widetilde{\chi}^2$($\dot{M}$, $T_{\rm eff,wd}$)
plane requires justification 
At issue is the question ``How dense a grid is adequate?" Phrased differently, ``As we produce
a denser and denser grid, and keeping in mind the committment of computer time, how do we know when to
stop?" A very important point is that our solution adopts values for the parameters M(wd), M(sec), Period,
$i$, parallax, and $E(B-V)$. A solution with a different value of any one of these would require recalculation
of an entire new plane of $\widetilde{\chi}^2$($\dot{M}$, $T_{\rm eff,wd}$) values, like Figure~5, and search of it for the 
minimum	$\widetilde{\chi}^2$. That process would be required, for example, if we study the implications of
changing the adopted parallax by one standard error. The solution we have found from our set of $\widetilde{\chi}^2$
values is valid only under the assumption that the other fixed parameter values are accurate.

Our Table~7 values for a specific $\dot{M}$ and variable $T_{\rm eff,wd}$, or specific $T_{\rm eff,wd}$ 
and variable $\dot{M}$, extend to a $\widetilde{\chi}^2$, 
when rounded to
the nearest half integer, of at least 4.0.
We consider that value to be enough larger than the minimum of
2.65, and the resolution in $\dot{M}$ and $T_{\rm eff,wd}$ sufficiently fine, 
to provide reasonable assurance that the (ideal) global minimum has been encircled. 
Figure~5, produced with IDL routine CONTOUR, presents contours of constant $\widetilde{\chi}^2$
including closed contours that encircle the tabular minimum.

We reiterate the point that possibly our most interesting
result is the accurate fit to a standard model. 
We note that a visual estimate ("chi by eye") selected the best model.

\citet{d2002} discuss the wind in V3885 Sgr and point to the lack of correlation between wind activity and
system luminosity, an inconsistency with expectation for line-driven winds. The theory of line-driven
winds has been applied to CV systems by \citet{pereyra2006} and \citet{proga2003a,proga2003b}.
See the review by \citet{drew2000}.
\citet{proga2003a} considers both radiation-driven and MHD effects and shows that hybrid models that include 
both MHD and
line-driving produce larger $\dot{M}_{\rm wind}$ values, of order $10^{-10}M_{\odot}~{\rm yr}^{-1}$,
but \citet{proga2003b} shows that a pure line-driven wind produces an improved fit to observed line
shapes but with a lower $\dot{M}_{\rm wind}$. As \citet{d2002} point out,
the lack of correlation between wind activity and system luminosity may argue in favor of MHD effects.
In either case, the theoretically derived values of $\dot{M}_{\rm wind}$ are far smaller than the mass
transfer $\dot{M}$ and essentially all of the mass transfer stream must be deposited on the WD.

Using eq.~2 of \citet{sion1995}, the $T_{\rm eff}$ of a compressionally-heated WD is given by
$T_{\rm eff} = [(fGM_{\rm wd}{\dot{M}})/(4{\pi}R_{\rm wd}^3{\sigma})]^{0.25}$, where $f$ is the
fraction of the accretion energy applied to compressional heating and is in the range 0.15-0.20.
For $f=0.15$ and the parameters of Table~8 we calculate $T_{\rm eff}$=52,500K, 
while for $f=0.20$ we find $T_{\rm eff}$=56,414K,
well within the estimated error limit of our 57,000K WD.

This study raises an obvious question: Why does a standard model accretion disk represent V3885 Sgr
accurately while other apparently comparable systems violate the standard model? To rephrase the
question: What equation or equations defining the standard model do the nonconforming systems violate?
The standard model equations (e.g., FKR eq. 5.68) adopt axial symmetry and
hydrostatic equilibrium, avoid MHD considerations (the MRI \citep{bh1991} source
of viscosity), and assume a constant $\alpha$ for the entire accretion disk, corresponding to a
constant local $\dot{M}$ within the accretion disk. The observed presence of 
spiral waves in V3885 Sgr \citep{Hartley2005}
violates the axial symmetry condition even in a system that fits the standard model. The impact of
the mass transfer stream produces a departure from axial symmetry. 
These phenomena illustrate the point that the standard model is a simplified but very useful idealization.
Just as weather in the Earth's
atmosphere, from physically complex phenomena, produces transient but extended local departures from 
the ideally accurate hydrostatic equilibrium, we speculate 
that transients
of unknown but physically complex origin
may occur in localized regions of accretion disks that perturb the radiation characteristics 
from the standard model average. 
$IUE$ spectra of UX~Ursae Majoris show substantial time-dependent SED changes for that NL system 
\citep{linnell2008b}. The adherence of V3885 Sgr to the standard model in this study may indicate an
infrequently violated average. 
On the other hand, the observation times of nonconforming systems may occur at times when transients are 
prevalent but detailed frequent monitoring might discover
times when their SEDs fit the standard model.

\section{Summary}

We use the \citet{knigge2007} period-secondary mass relation to assign the secondary mass and the mean WD mass
from the same paper, consistent with an observationally-determined mass ratio, to set the V3885 Sgr WD mass.
Based on a $\widetilde{\chi}^2$ analysis using 38 system models and corresponding synthetic spectra,
we find that a standard model accretion disk with $\dot{M}=5.0{\pm}2.0{\times}10^{-9}~M_{\odot}~{\rm yr}^{-1}$,
a $57,000{\pm}5000$K $0.7M_{\odot}$ WD, and an orbital inclination of $65{\pm}2{\arcdeg}$ provide 
a system synthetic spectrum which optimally fits observed 
$FUSE$ and STIS
spectra of V3885 Sgr. The result is robust for small changes in the assumed $M_{\rm wd}$.
A $Hipparcos$	parallax fixes the divisor enabling the system synthetic spectrum to be
matched to the observed spectra and establishes the fit on an absolute flux basis.
The model thus provides realistic constraints on $\dot{M}$ and $T_{\rm eff}$ for a large $\dot{M}$
system above the period gap.
%The derived $\dot{M}$ represents a reliable value within a specified range for a system above the period gap, 
%and the derived $T_{\rm eff,wd}$ represents a constrained value for a large $\dot{M}$
%system.

We are grateful to the referee for a careful reading of the paper and for a series of comments that
led to a significant improvement of the paper.
P.S. acknowledges support from $FUSE$ grant NNG04GC97G and HST grant GO-09724. 
Support for this work was provided by NASA through grant number 
HST-AR-10657.01-A  to Villanova University (P. Godon) from the Space
Telescope Science Institute, which is operated by the Association of
Universities for Research in Astronomy, Incorporated, under NASA
contact NAS5-26555. Additional support for this work was provided by the National Aeronautics
and Space Administration (NASA) under Grant number NNX08AJ39G issued through the office
of Astrophysics Data Analysis Program (ADP) to Villanova University (P. Godon).
Participation by E.~M.~Sion, P.~Godon and A. Linnell was also supported in part by NSF grant AST0807892
to Villanova University.
This research was partly based on observations made with
the NASA-CNES-CSA Far Ultraviolet Spectroscopic Explorer. $FUSE$ is operated
for NASA by the Johns Hopkins University under NASA contract NAS5-32958.

\clearpage

%% Use the figure environment and \plotone or \plottwo to include 
%% figures and captions in your electronic submission.

%%%%%%%%%%%%%%%%%%%%%%%%%%%%%%%%%%%%%%%%%%%%%%%%%%%%%%%%%%%%%%%%%%%
%%% TABLES
%%%%%%%%%%%%%%%%%%%%%%%%%%%%%%%%%%%%%%%%%%%%%%%%%%%%%%%%%%%%%%%%%%%

\clearpage

%%%%%%%%%%%%%%%%%%%%%%%%%%%%%%%%%%%%%%%%%%%%%%%%%%%%%%%%%%%%%%%%%%%
\setlength{\hoffset}{-30mm}
\begin{table} 
\caption{FUV Observations of V3885 Sgr: HST \& FUSE Spectra} 
\begin{tabular}{clcllll}
\hline
 Date      & Telescope   & Exp time & Dataset   & Filter/Grating &Operation & Calibration  \\ 
(dd/mm/yyyy) & /Instrument & (sec)  &           & /Aperture      & Mode     & Software \\ 
\hline
 24 May 2000 & FUSE        & 12392     & P1870101000  & LWRS           & TTAG     & CalFUSE 3.0.7 \\ 
 30 Apr 2000 & STIS        & 2195     & O5BI04010 & E140M/0.2x0.2 & TIME-TAG & CALSTIS 2.22 \\  
\hline
\end{tabular}
\end{table}
\setlength{\hoffset}{00mm}

%%%%%%%%%%%%%%%%%%%%%%%%%%%%%%%%%%%%%%%%%%%%%%%%%%%%%%%%%%%%%%%%%%%

\clearpage

%%%%%%%%%%%%%%%%%%%%%%%%%%%%%%%%%%%%%%%%%%%%%%%%%%%%%%%%%%%%%%%%%%%
\begin{deluxetable}{llll}
\tablewidth{0pt}
\tablenum{2}
\tablecaption{FUSE lines
}
\tablehead{
\colhead{Line}  &  \colhead{Wavelength} & \colhead{$\lambda_0$} 
& \colhead{Origin}\\	
\colhead{Identification}	& \colhead{(\AA)}	& \colhead{(\AA)} \\
}
\startdata
H\,{\sc i}    & 914.13      &      914.29 &   ism   \\         
H\,{\sc i}    & 914.43      &      914.58 &   ism   \\         
H\,{\sc i}    & 914.79      &      914.92 &   ism   \\         
H\,{\sc i}    & 915.19      &      915.33 &   ism   \\         
H\,{\sc i}    & 915.68      &      915.82 &   ism   \\         
H\,{\sc i}    & 916.30      &      916.43 &   ism   \\         
O\,{\sc i}    & 916.73      &      916.82 &   ism   \\         
H\,{\sc i}    & 917.05      &      917.18 &   ism   \\         
H\,{\sc i}    & 918.00      &      918.13 &   ism   \\         
H\,{\sc i}    & 919.22      &      919.35 &   ism   \\         
H\,{\sc i}    & 920.84      &      920.96 &   ism   \\         
H\,{\sc i}    & 921.75      &      921.86 &   ism   \\         
H\,{\sc i}    & 923.01      &      923.15 &   ism   \\         
N\,{\sc iv}&$\sim$922       &      923.2  &   s   \\         
O\,{\sc i}    & 924.83      &      924.95 &   ism   \\         
O\,{\sc i}    & 925.33      &      925.45 &   ism   \\         
H\,{\sc i}    & 926.11      &      926.23 &   ism   \\         
O\,{\sc i}    & 929.41      &      929.52 &   ism   \\         
O\,{\sc i}    & 930.13      &      930.26 &   ism   \\         
H\,{\sc i}    & 930.61      &      930.75 &   ism   \\         
S\,{\sc vi}&$\sim$932       &      933.38 &     s     \\         
H\,{\sc i}    & 936.51      &      936.75 &   ism   \\         
H\,{\sc i}    & 937.67      &      937.80 &   ism   \\         
S\,{\sc vi}&$\sim$943       &      944.52 &     s     \\         
O\,{\sc i}    & 948.57      &      948.69 &   ism   \\         
H\,{\sc i}    & 949.61      &      949.74 &   ism   \\         
O\,{\sc i}    & 950.76      &      950.88 &   ism   \\         
N\,{\sc i}+O\,{\sc i}   & 952.26      &             &   ism   \\         
N\,{\sc i}    & 953.30      &      953.42 &   ism     \\         
N\,{\sc i}    & 953.54      &      953.65 &   ism     \\         
N\,{\sc i}    & 953.87      &      953.97 &   ism     \\         
N\,{\sc i}    & 953.97      &      954.10 &   ism     \\         
P\,{\sc ii}   & 962.44      &      962.57 &   ism     \\         
P\,{\sc ii}   & 963.68      &      963.80 &   ism     \\         
N\,{\sc i}    & 963.87      &      963.99 &   ism     \\         
N\,{\sc i}    & 964.50      &      964.63 &   ism     \\         
N\,{\sc i}    & 964.94      &      965.04 &   ism     \\ 
O\,{\sc i}    & 971.61      &      971.74 &   ism   \\         
H\,{\sc i}    & 972.34      &      972.54 &   ism   \\         
C\,{\sc iii}  & $\sim$976   &      977.02 &     s      \\
C\,{\sc iii}  & 976.88      &      977.02 &   ism     \\         
O\,{\sc i}    & 976.32      &      976.45 &   ism   \\         
O\,{\sc i}    & 988.56      &      988.65 &   ism   \\         
Si\,{\sc ii}  & 989.75      &      989.87 &   ism   \\         
N\,{\sc iii}  & $\sim$990   &      991.5  &    s      \\
Si\,{\sc iii} & 997.26      &      997.38 &   ism     \\         
Si\,{\sc ii}  & 1020.63     &     1020.70 &   ism   \\         
H\,{\sc i}    & 1025.37     &     1025.72 &   s,ism \\ 
H\,{\sc i}    & 1025.68     &     1025.72 &   s,ism \\ 
O\,{\sc vi}   & 1030.7      &     1031.93 &   s       \\ 
O\,{\sc vi}   & 1036.4      &     1037.62 &   s       \\ 
C\,{\sc ii}   & 1036.24     &     1036.34 &   ism     \\
O\,{\sc i}    & 1039.18     &     1039.23 &   ism    \\ 
Ar\,{\sc i}   & 1048.15     &     1048.22  &   ism     \\ 
S\,{\sc iv}   & 1062.3      &     1062.65  &    s      \\
Ar\,{\sc i}   & 1066.59     &     1066.66  &  ism     \\ 
S\,{\sc iv}   & 1072.5      &     1072.97  &    s      \\ 
N\,{\sc ii}   & 1083.74     &     1083.99  &    ism   \\
Si\,{\sc iii} & 1109.4      &     1108.36  &   s   \\  
              &             &     1109.94  &   s   \\  
Si\,{\sc iii} & 1113.0      &     1113.2   &   s   \\  
S\,{\sc iv}   & 1117.3      &     1117.76  &   s   \\ 
P\,{\sc v}    & 1117.3      &     1117.98  &   s   \\ 
              & 1127.4      &     1128.01  &   s   \\ 
Si\,{\sc iv}  & 1122.0      &     1122.5   &   s   \\ 
              & 1127.4      &     1128.3   &   s   \\
N\,{\sc i}    & 1134.08     &     1134.16  &   ism       \\
N\,{\sc i}    & 1134.33     &     1134.42  &   ism     \\
N\,{\sc i}    & 1134.89     &     1134.98  &   ism     \\
C\,{\sc iii}  & 1175.0      &1174.9-1176.4 &   s       \\

\enddata
\tablecomments{a=possibly with terrestrial contamination; 
ism=ISM; s=source}
\end{deluxetable}

%%%%%%%%%%%%%%%%%%%%%%%%%%%%%%%%%%%%%%%%%%%%%%%%%%%%%%%%%%%%%%%%%%%

%%%%%%%%%%%%%%%%%%%%%%%%%%%%%%%%%%%%%%%%%%%%%%%%%%%%%%%%%%%%%%%%%%%

\begin{deluxetable}{llll}
\tablewidth{0pt}
\tablenum{3}
\tablecaption{STIS lines
}
\tablehead{
\colhead{Line}  &  \colhead{Wavelength} & \colhead{$\lambda_0$} 
& \colhead{Origin}\\	
\colhead{Identification}	& \colhead{(\AA)}	& \colhead{(\AA)} \\
}
\startdata
C\,{\sc iii}  & 1173.62       & 1174.9-1176.4 &   s        \\
Si\,{\sc ii}  & 1190.36       & 1190.42   &  ism         \\
              & 1193.24       & 1193.29   &  ism       \\
N\,{\sc i}    & 1199.49       & 1199.55   &  ism   \\  
              & 1200.18       & 1200.22   &  ism   \\  
              & 1200.67       & 1200.71   &  ism   \\  
Si\,{\sc iii} & 1204.77       & 1206.50   &     s      \\
Si\,{\sc iii} & 1206.45       & 1206.50   &  ism       \\
H\,{\sc i}    & 1215.6        & 1215.67   &    s       \\
N\,{\sc v}    & 1236.9        & 1238.82 &   s         \\ 
              & 1240.3        & 1242.80 &   s         \\
N\,{\sc v}    & 1238.63       & 1238.82 &   ism         \\ 
              & 1242.59       & 1242.80 &   ism         \\
S\,{\sc ii}   & 1250.55       & 1250.58  &  ism \\ 
              & 1253.77       & 1253.81  &   ism     \\ 
              & 1259.45       & 1259.52  &   ism \\ 
Si\,{\sc ii}  & 1260.37       & 1260.42  &   ism \\ 
Si\,{\sc iii} & 1297.3        & 1294.55  &   s   \\ 
              &               & 1296.73  &   s   \\ 
              &               & 1298.89  &   s   \\ 
O\,{\sc i}    & 1302.12       & 1302.17 &   ism   \\ 
Si\,{\sc ii}  & 1304.33       & 1304.37 &   ism   \\ 
C\,{\sc ii}   & 1334.47       & 1334.53 &   ism    \\ 
              & 1334.47       & 1334.66 &   ism    \\ 
              & 1335.66       & 1335.71 &   ism     \\
Si\,{\sc iv}  & 1391.4        & 1393.76      &   s        \\
              & 1400.35       & 1402.77      &   s        \\
Si\,{\sc ii}  & 1526.62       & 1526.71      &  ism        \\
C\,{\sc iv}   &$\sim$1545     & 1548.20       &   s        \\
              &               & 1550.78       &   s        \\
C\,{\sc iv}   & 1547.95       & 1548.20       &  ism        \\
              & 1550.54       & 1550.78       &  ism       \\
He\,{\sc ii}  & 1637.7        & 1640.5        &   s        \\
Al\,{\sc ii}  & 1670.73       & 1670.8        &  ism    \\
\enddata
\tablecomments{ 
ism=ISM; s=source;
there is possibly some atmospheric absorption contamination
for N\,{\sc i}, N\,{\sc ii}, H\,{\sc i}, O\,{\sc i}.  
}
\end{deluxetable}

%%%%%%%%%%%%%%%%%%%%%%%%%%%%%%%%%%%%%%%%%%%%%%%%%%%%%%%%%%%%%%%%%%%

\clearpage

%%%%%%%%%%%%%%%%%%%%%%%%%%%%%%%%%%%%%%%%%%%%%%%%%%%%%%%%%%%%%%%%%%%
\begin{deluxetable}{llll}
\tablewidth{0pt}
\tablenum{4}
\tablecaption{V3885 Sgr Initial System Parameters}
\tablehead{
\colhead{parameter} & \colhead{value} & \colhead{parameter} & \colhead{value}}
\startdata
$M_{\rm wd}$  &  $0.70M_{\odot}$\tablenotemark{a}	 & $i$    &  $65{\arcdeg}$~\tablenotemark{d}\\
${M}_2$  &  $0.475{M}_{\odot}$\tablenotemark{b}	 & ${\dot{M}}$ 	&$5.0{\times}10^{-9}{M}_{\odot} 
{\rm yr}^{-1}$~\tablenotemark{e} \\
P    &  0.20716071 day~\tablenotemark{c}	  & $d$     	& $110{\pm}30$~pc~\tablenotemark{f}\\
\enddata
\tablenotetext{a}{\citet{knigge2007} mean $M_{\rm wd}$; \citet{Hartley2005}, $0.55M_{\odot}$ to 
$0.8M_{\odot}$;
\citet{Ribiero2007}, $0.3M_{\odot}$ to $0.9M_{\odot}$}
\tablenotetext{b}{\citet{knigge2007} calibrated $P:M_2$ relation; \citet{Hartley2005}, 
$q$=0.6 to 0.8; \citet{Ribiero2007}, $M_2=0.2M_{\odot}$ to $0.55M_{\odot}$}
\tablenotetext{c}{\citet{Ribiero2007}}
\tablenotetext{d}{\citet{Haug1985}, $i$ between $60{\arcdeg}$ and $70{\arcdeg}$;
\citet{Ribiero2007}, $i$ between $45{\arcdeg}$ and $75{\arcdeg}$;
\citet{Hartley2005}, $i>65{\arcdeg}$}
\tablenotetext{e}{\citet{linnell2007} and similarity to IX Vel (\citet{d2002})}
\tablenotetext{f}{Hipparcos value}
\end{deluxetable}

%%%%%%%%%%%%%%%%%%%%%%%%%%%%%%%%%%%%%%%%%%%%%%%%%%%%%%%%%%%%%%%%%%%
\clearpage
%%%%%%%%%%%%%%%%%%%%%%%%%%%%%%%%%%%%%%%%%%%%%%%%%%%%%%%%%%%%%%%%%%%

\begin{deluxetable}{rrrrrrrr}
\tablewidth{0pt}
\tablenum{5}
\tablecaption{Properties of accretion disk with mass transfer rate 
$\dot{M}=5.0{\times}10^{-9}~{M}_{\odot}{\rm yr}^{-1}$ and WD mass of $0.70{M}_{\odot}$.}
\tablehead{	  
\colhead{$r/r_{\rm wd,0}$} & \colhead{$T_{\rm eff}$} & \colhead{$m_0$} 
& \colhead{log~$g$}
& \colhead{$z_0$} & \colhead{$Ne$} & \colhead{{$\tau_{\rm Ross}$}}
}
\startdata
1.36  &  53136  &  1.221E4   &  6.84   & 8.27E7  & 2.0E17  & 2.33E4\\
2.00  &	 47642  &  1.491E4	 &  6.58   & 1.43E8  & 1.3E17  & 2.78E4\\
3.00  &	 38525  &  1.456E4	 &  6.28   & 2.43E8  & 7.7E16  & 3.35E4\\
4.00  &	 32380  &  1.356E4	 &  6.06   & 3.49E8  & 5.1E16  & 3.92E4\\
5.00  &	 28086  &  1.259E4	 &  5.89   & 4.60E8  & 3.8E16  & 4.52E4\\
6.00  &	 24918  &  1.176E4	 &  5.75   & 5.76E8  & 3.0E16  & 5.16E4\\
8.00  &	 20531  &  1.043E4	 &  5.53   & 8.14E8  & 2.5E16  & 6.56E4\\
10.00 &	 17612  &  9.439E3	 &  5.35   & 1.06E9  & 1.9E16  & 8.05E4\\
12.00 &	 15512  &  8.664E3	 &  5.21   & 1.32E9  & 1.5E16  & 9.53E4\\
14.00 &	 13923  &  8.040E3	 &  5.09   & 1.58E9  & 1.2E16  & 1.08E5\\
16.00 &	 12670  &  7.526E3	 &  4.98   & 1.85E9  & 9.9E15  & 1.18E5\\
18.00 &	 11653  &  7.092E3	 &  4.89   & 2.12E9  & 8.3E15  & 1.23E5\\
20.00 &	 10810  &  6.721E3	 &  4.80   & 2.38E9  & 6.7E15  & 1.23E5\\
24.00 &	  9487  &  6.115E3	 &  4.65   & 2.92E9  & 5.1E15  & 1.17E5\\
28.00 &	  8491  &  5.638E3	 &  4.51   & 3.57E9  & 4.7E15  & 1.07E5\\
32.00 &	  7711  &  5.251E3	 &  4.46   & 4.90E9  & 2.2E14  & 9.32E4\\
\enddata
\tablecomments{Each line in the table represents a separate annulus.
A \citet{ss1973} viscosity parameter $\alpha=0.1$ was used in calculating all annuli.
The WD radius, $r_{\rm wd,0}$, is the radius, $0.01091R_{\odot}$, of a zero temperature 
Hamada-Salpeter carbon model. See the text (\S4.) for a discussion of the table units.
}		 
\end{deluxetable}

%%%%%%%%%%%%%%%%%%%%%%%%%%%%%%%%%%%%%%%%%%%%%%%%%%%%%%%%%%%%%%%%%%%
\clearpage
%%%%%%%%%%%%%%%%%%%%%%%%%%%%%%%%%%%%%%%%%%%%%%%%%%%%%%%%%%%%%%%%%%%
\begin{deluxetable}{rrrrrr}
\tablewidth{0pt}
\tablenum{6}
\tablecaption{Temperature profile for V3885 Sgr 
accretion disk with mass transfer rate of
$\dot{M}=5.0{\times}10^{-9}~{M}_{\odot}{\rm yr}^{-1}$ and WD mass of $0.70{M}_{\odot}$.}
\tablehead{	  
\colhead{$r/r_{\rm wd}$} & \colhead{$T_{\rm eff}$} & \colhead{$r/r_{\rm wd}$} & \colhead{$T_{\rm eff}$} 
& \colhead{$r/r_{\rm wd}$} & \colhead{$T_{\rm eff}$}
}
\startdata
1.00    &	43608	&	13.32	&	12308	&	27.94	&	7260\\
1.18	&	43608	&	14.65	&	11511	&	29.27	&	7021\\
1.36	&	45358	&	15.98	&	10826	&	30.59	&	6799\\
2.69	&	34985	&	17.31	&	10229	&	31.92	&	6594\\
4.02	&	27556	&	18.64	&	9705	&	33.25	&	6402\\
5.35	&	22948	&	19.96	&	9240	&	34.58	&	6223\\
6.68	&	19807   &	21.29	&	8824	&	35.91	&	6055\\
8.01	&	17518	&	22.62	&	8449	&	37.24	&	5898\\
9.33	&	15767	&	23.95	&	8110	&	38.57	&	5749\\
10.66	&	14380	&	25.28	&	7802	&	39.90	&	5610\\
11.99	&	13249	&	26.61	&	7520	&	41.22	&	5478\\
\enddata
\tablecomments{
The WD radius $r_{\rm wd}$ is the radius of a 60,000K $0.70{M}_{\odot}$ WD interpolated from
Table 4a of \citet{panei2000}. The $T_{\rm eff}=43608$K for $r/r_{\rm wd}$=1.00 refers to
the inner edge of the innermost annulus. The same $T_{\rm eff}$ for $r/r_{\rm wd}$=1.18
refers to the outer edge of the innermost annulus which coincides with the inner edge of
the next annulus. The remaining $T_{\rm eff}$ values refer to the inner edge of the
corresponding annuli.
}		 
\end{deluxetable}

%%%%%%%%%%%%%%%%%%%%%%%%%%%%%%%%%%%%%%%%%%%%%%%%%%%%%%%%%%%%%%%%%%%

\clearpage

%%%%%%%%%%%%%%%%%%%%%%%%%%%%%%%%%%%%%%%%%%%%%%%%%%%%%%%%%%%%%%%%%%%
\begin{deluxetable}{cccrcccr}
\tablewidth{0pt}
\tablenum{7} 
\tablecaption{Values of $\widetilde{\chi}^2$ as function of model parameters}
\tablehead{ 
\colhead{$\dot{M}$ (x$10^{-9} M_{\odot}/{\rm{yr}}$)}  & \colhead{$T_{\rm{wd}}$(K)}  & \colhead{$i$(deg.)}   
& \colhead{$\widetilde{\chi}^2$} & \colhead{$\dot{M}$ (x$10^{-9} M_{\odot}/{\rm{yr}}$)}  & \colhead{$T_{\rm{wd}}$(K)}  
& \colhead{$i$(deg.)} & \colhead{$\widetilde{\chi}^2$}
}    
\startdata
 4.0	&		70,000		&	65	&			7.51	&   5.0 &		57,000		&	64	&	3.95\\
 4.0	&		75,000		&	65	&			5.60	&	5.0	&		57,000		&	66	&	4.43\\
 4.5	&		57,000		&	65	&			6.86	&	5.0 &		57,000		&	70	&  39.73\\	
 4.5	&		60,000		&	65	&			5.05	& 	5.2 &		40,000		&	65	&	4.69\\
 4.5	&		65,000		&	65	&			3.93	&   5.2	&		45,000		&	65	&	3.14\\  
 4.5	&		70,000		&	65	&			3.14	&   5.2	&		50,000		&	65	&	2.91\\
 4.5	&		75,000		&	65	&			3.97	&   5.2	&		52,000		&	65	&	2.73\\
 4.8	&		52,000		&	65	&			5.22	&   5.2	&		57,000		&	65	&	3.04\\
 4.8	&		57,000		&	65	&			3.42	&   5.2	&		60,000		&	65	&	4.41\\
 4.8	&		60,000		&	65	&			3.00	&   5.5	&		35,000		&	65	&	4.34\\
 4.8	&		65,000		&	65	&			3.40	&   5.5	&		40,000		&	65	&	3.08\\
 4.8	&		70,000		&	65	&			4.26	&   5.5	&		45,000		&	65	&	3.19\\
 5.0	&		45,000		&	65	&			4.54	&   5.5	&		50,000		&	65	&	3.24\\
 5.0	&		50,000		&	65	&			4.25	&   5.5	&		52,000		&	65	&	3.93\\
 5.0	&		52,000		&	65	&			3.39	&   5.5	&		57,000		&	65	&	5.93\\
 5.0	&		57,000		&	65	&			2.65	&   6.0	&		35,000		&	65	&	4.03\\
 5.0	&		60,000		&	65	&			3.44	&   6.0	&		40,000		&	65	&	6.48\\
 5.0	&		65,000		&	65	&			5.80	&   6.0	&		45,000		&	65	&	4.13\\
 5.0	&		57,000		&	60	&		   49.94	&   6.0	&		50,000		&	65	&	9.88\\
 \enddata
\end{deluxetable}

%%%%%%%%%%%%%%%%%%%%%%%%%%%%%%%%%%%%%%%%%%%%%%%%%%%%%%%%%%%%%%%%%%%
\clearpage
%%%%%%%%%%%%%%%%%%%%%%%%%%%%%%%%%%%%%%%%%%%%%%%%%%%%%%%%%%%%%%%%%%%
\begin{deluxetable}{llll}
\tablewidth{0pt}
\tablenum{8}
\tablecaption{V3885 Sgr Model System Parameters}
\tablehead{
\colhead{parameter} & \colhead{value} & \colhead{parameter} & \colhead{value}}
\startdata
${ M}_{\rm wd}$  &  $0.70{\pm}0.1{M}_{\odot}$	 & $T_{\rm eff,s}$(pole)    &  $3650{\pm}100$K \\
${M}_{\rm s}$  &  $0.475{M}_{\odot}$	 & $T_{\rm eff,s}$(point)	&	2214K\\
${\dot{M}}$      &  $5.0{\pm}2.0{\times}10^{-9}{M}_{\odot} {\rm yr}^{-1}$ &	 $T_{\rm eff,s}$(side)	&	3598K\\
P    &  0.2071607 days	  &  	 $T_{\rm eff,s}$(back)	&	3489K \\
$D$              &  $1.55396R_{\odot}$ & $r_s$(pole) &  $0.209R_{\odot}$\\	 
${\Omega}_{\rm wd}$         & 116.0 & $r_s$(point)   & $0.296R_{\odot}$\\
${\Omega}_s$                &  3.2053112 & $r_s$(side)  & $0.218R_{\odot}$\\
{\it i}              &   $65{\pm}2{\degr}$& $r_s$back)   & $0.239R_{\odot}$\\
$T_{\rm eff,wd}$         &  $57,000{\pm}5000$K  & log $g_s$(pole) & 4.73\\
$r_{\rm wd}$      &   $0.0134R_{\odot}$ & log $g_s$(point) & -.390\\
log $g_{\rm wd}$  &   8.02& log $g_s$(side)  & 4.65\\
$A_{\rm wd}$        &  1.0 & log $g_s$(back)  & 4.48\\ 
$A_s$               &  0.6  & $r_a$ & $0.56R_{\odot}$\\
${\beta}_{\rm wd}$  &  0.25 & $r_b$ & $0.01091R_{\odot}$\\
${\beta}_s$         &  0.08& $H$    & $0.10R_{\odot}$\\
\enddata
\tablecomments{${\rm wd}$ refers to the WD; $s$ refers to the secondary star.
$D$ is the component separation of centers,
${\Omega}$ is a Roche potential. Temperatures are polar values, 
$A$ values are bolometric albedos, and $\beta$ values are 
gravity-darkening exponents. 
$r_a$ specifies the outer radius 
of the accretion disk, set at the tidal cut-off radius, 
and $r_b$ is the accretion disk inner radius, as 
determined in the final system model, while $H$ is 
the semi-height of the accretion disk rim.}  
\end{deluxetable}

%%%%%%%%%%%%%%%%%%%%%%%%%%%%%%%%%%%%%%%%%%%%%%%%%%%%%%%%%%%%%%%%%%%
\clearpage

%%%%%%%%%%%%%%%%%%%%%%%%%%%%%%%%%%%%%%%%%%%%%%%%%%%%%%%%%%%%%%%%%%%

\begin{deluxetable}{ll}
\tablewidth{0pt}
\tablenum{9} 
\tablecaption{Parameters of ISM model}
\tablehead{ 
\colhead{Parameter}           & \colhead{Value}}     
\startdata
N(H)                & $1.0{\times}10^{18} {\rm cm}^{-2}$\\
D/H                 & $2{\times}10^{-5}$\\
N(H2)               & $1{\times}10^{14} {\rm cm}^{-2}$\\
vel.                & $-50.0 {\rm~km~s^{-1}}$\\
temp.               & 250 K\\
metallicity         & 1.0 (1.0 = solar)\\
\enddata
\end{deluxetable}

%%%%%%%%%%%%%%%%%%%%%%%%%%%%%%%%%%%%%%%%%%%%%%%%%%%%%%%%%%%%%%%%%%%

%%%%%%%%%%%%%%%%%%%%%%%%%%%%%%%%%%%%%%%%%%%%%%%%%%%%%%%%%%%%%%%%%%%
%%% FIGURES
%%%%%%%%%%%%%%%%%%%%%%%%%%%%%%%%%%%%%%%%%%%%%%%%%%%%%%%%%%%%%%%%%%%

\clearpage

%%%%%%%%%%%%%%%%%%%%%%%%%%%%%%%%%%%%%%%%%%%%%%%%%%%%%%%%%%%%%%%%%%%
\begin{figure}[tb]
%\epsscale{0.97}
\includegraphics[scale=.80,angle=-90]{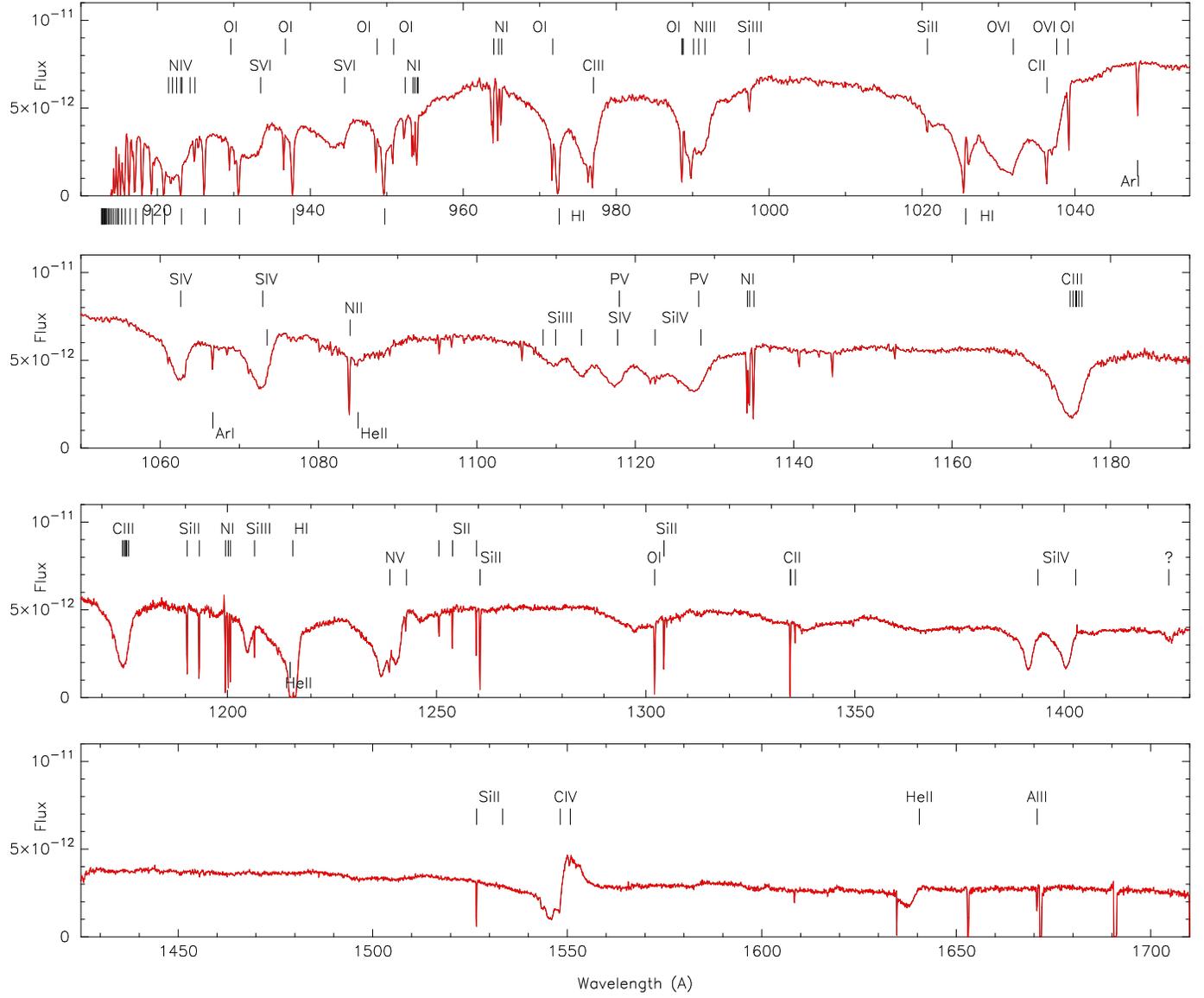}
%\plotfiddle{lines.ps}{0.80}{90}{500pt}{450pt}{pt}{0pt}
%\epsscale{0.80}
\figcaption{
$FUSE$ plus $STIS$ combined spectra with line identifications.
\label{fg1}}
\end{figure}
%%%%%%%%%%%%%%%%%%%%%%%%%%%%%%%%%%%%%%%%%%%%%%%%%%%%%%%%%%%%%%%%%%%

%%%%%%%%%%%%%%%%%%%%%%%%%%%%%%%%%%%%%%%%%%%%%%%%%%%%%%%%%%%%%%%%%%%
\begin{figure}[tb]
\epsscale{0.97}
\plotone{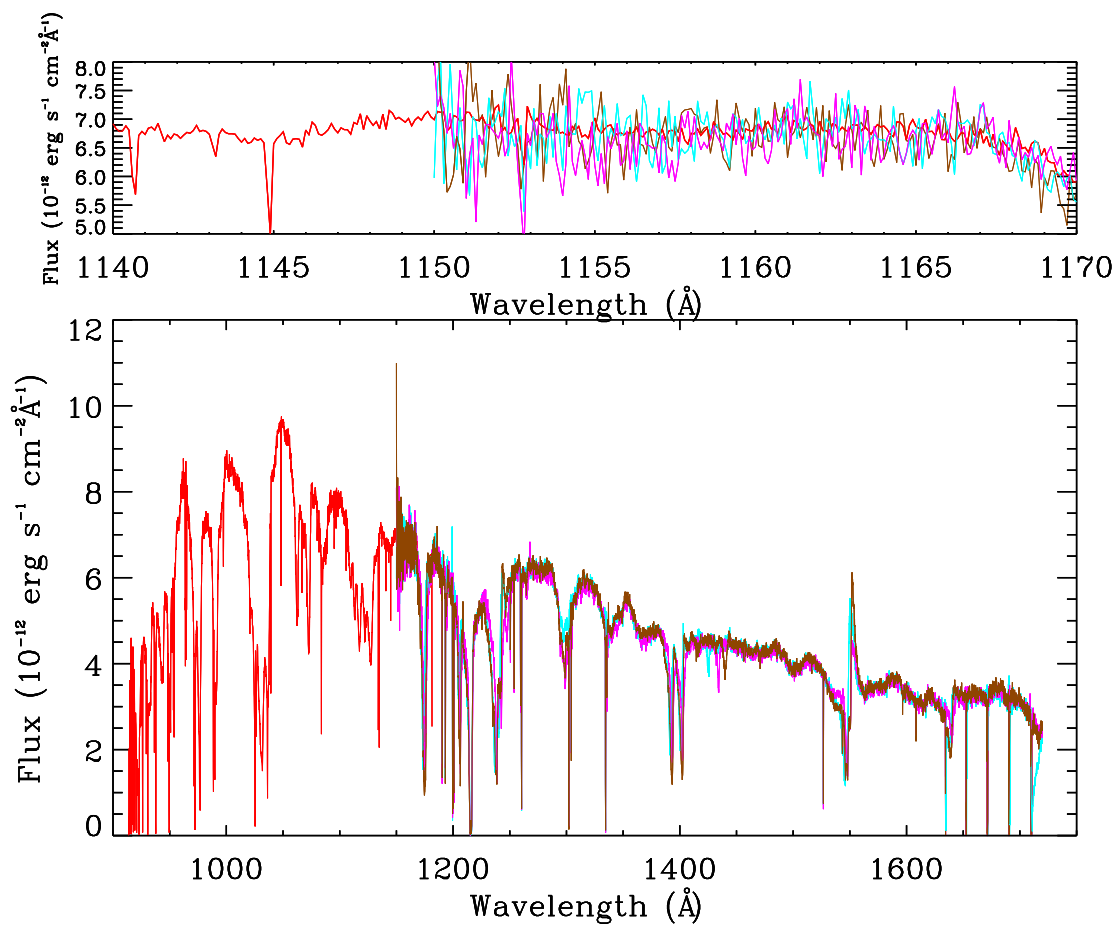}
\epsscale{1.00}
\vspace{5pt}
\figcaption{Combined $FUSE$ and STIS spectra. The upper panel shows the
fit of the spectra in the overlap region when small normalizing
factors are applied to the STIS spectra. The $FUSE$ spectrum is red;
the first STIS spectrum (Table~1), divided by 0.98, is cyan; the second
STIS spectrum, divided by 1.12, is magenta; the third STIS spectrum,
divided by 0.94, is brown. Note the accurate superposition of the
STIS spectra with application of normalizing factors.
\label{fg2}}
\end{figure}
%%%%%%%%%%%%%%%%%%%%%%%%%%%%%%%%%%%%%%%%%%%%%%%%%%%%%%%%%%%%%%%%%%%

%%%%%%%%%%%%%%%%%%%%%%%%%%%%%%%%%%%%%%%%%%%%%%%%%%%%%%%%%%%%%%%%%%%
\begin{figure}[tb]
\epsscale{0.97}
\plotone{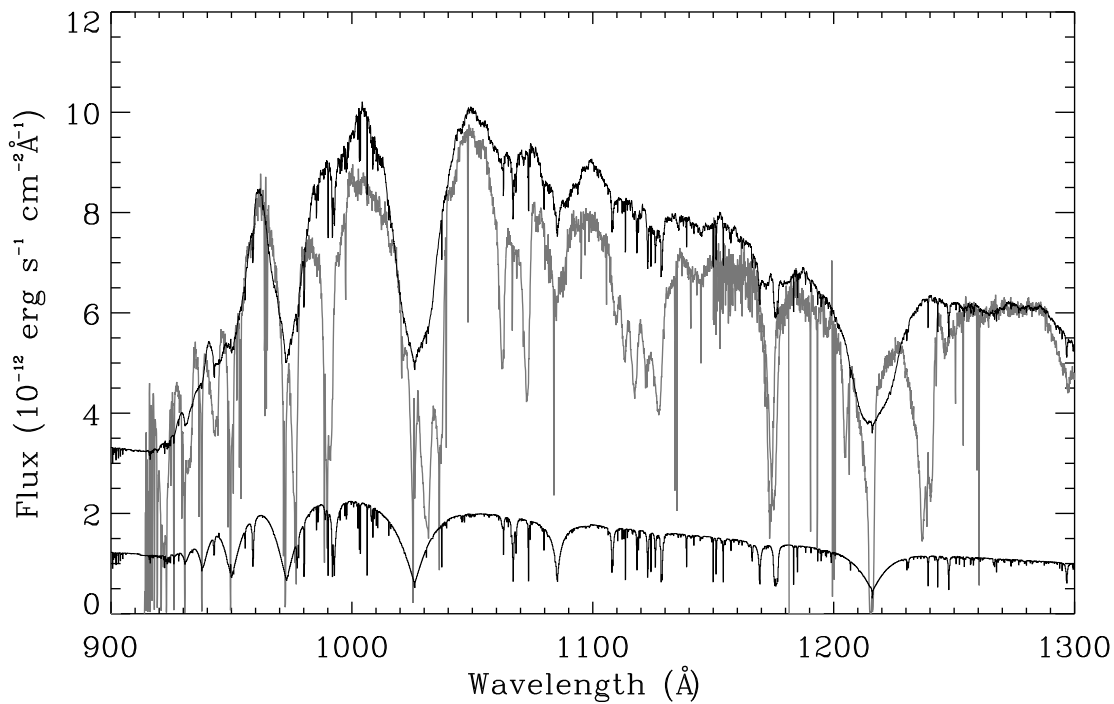}
\epsscale{1.00}
\vspace{5pt}
\figcaption{
FUV fit of system synthetic spectrum to $FUSE$ and STIS combined spectrum
(gray plot) of V3885Sgr. Note the extensive absorption by local highly
ionized material. The lower spectrum is the contribution of a 57,000K WD;
the top synthetic spectrum is the system synthetic spectrum including
the WD and a $\dot{M}=5.0{\times}10^{-9}~M_{\odot}~{\rm yr}^{-1}$
standard model accretion disk.
See the text for a discussion.
\label{fg3}}
\end{figure}
%%%%%%%%%%%%%%%%%%%%%%%%%%%%%%%%%%%%%%%%%%%%%%%%%%%%%%%%%%%%%%%%%%%

%%%%%%%%%%%%%%%%%%%%%%%%%%%%%%%%%%%%%%%%%%%%%%%%%%%%%%%%%%%%%%%%%%%
\begin{figure}[tb]
\epsscale{0.97}
\plotone{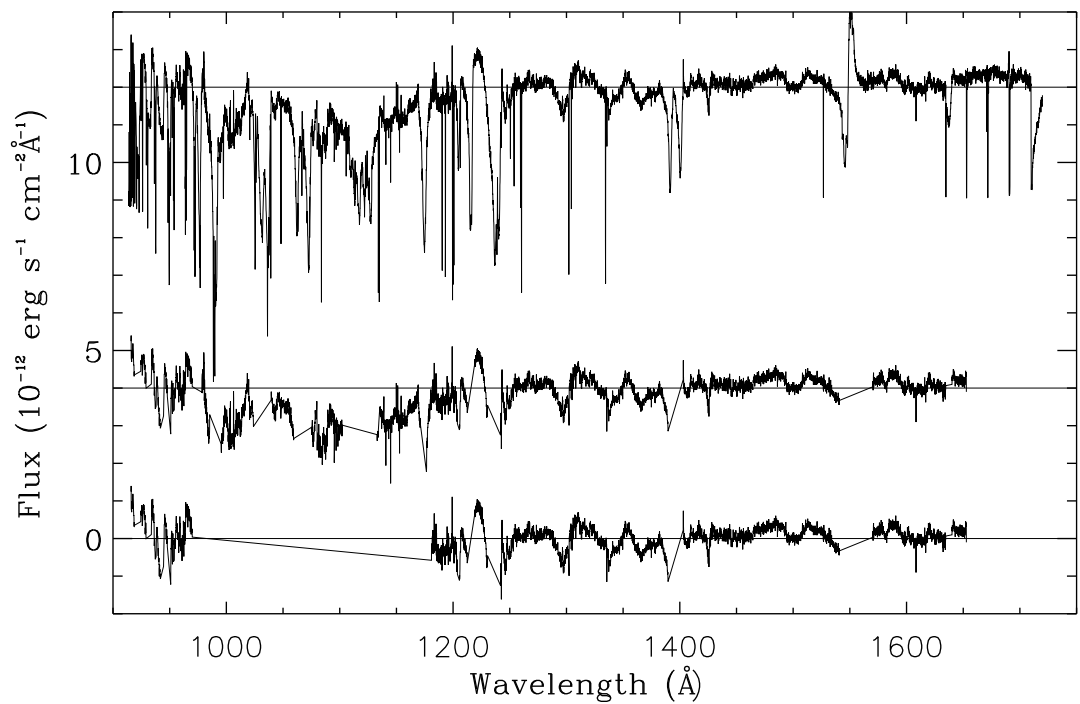}
\epsscale{1.00}
\vspace{5pt}
\figcaption{
Plots of residuals from the visually selected best model. The top plot,
displaced upward by 12.0 ordinate units, is the residuals in the
unmasked case.
The middle plot, displaced upward by 4.0 ordinate units, is the
residuals for a partially masked observed spectrum. The bottom plot
is the residuals for the finally adopted masked case.
Note the three horizontal lines marking residual values of 0.0 for
the three cases.
See the text for a discussion.
\label{fg4}}
\end{figure}
%%%%%%%%%%%%%%%%%%%%%%%%%%%%%%%%%%%%%%%%%%%%%%%%%%%%%%%%%%%%%%%%%%%

%%%%%%%%%%%%%%%%%%%%%%%%%%%%%%%%%%%%%%%%%%%%%%%%%%%%%%%%%%%%%%%%%%%
\begin{figure}[tb]
\epsscale{0.97}
\plotone{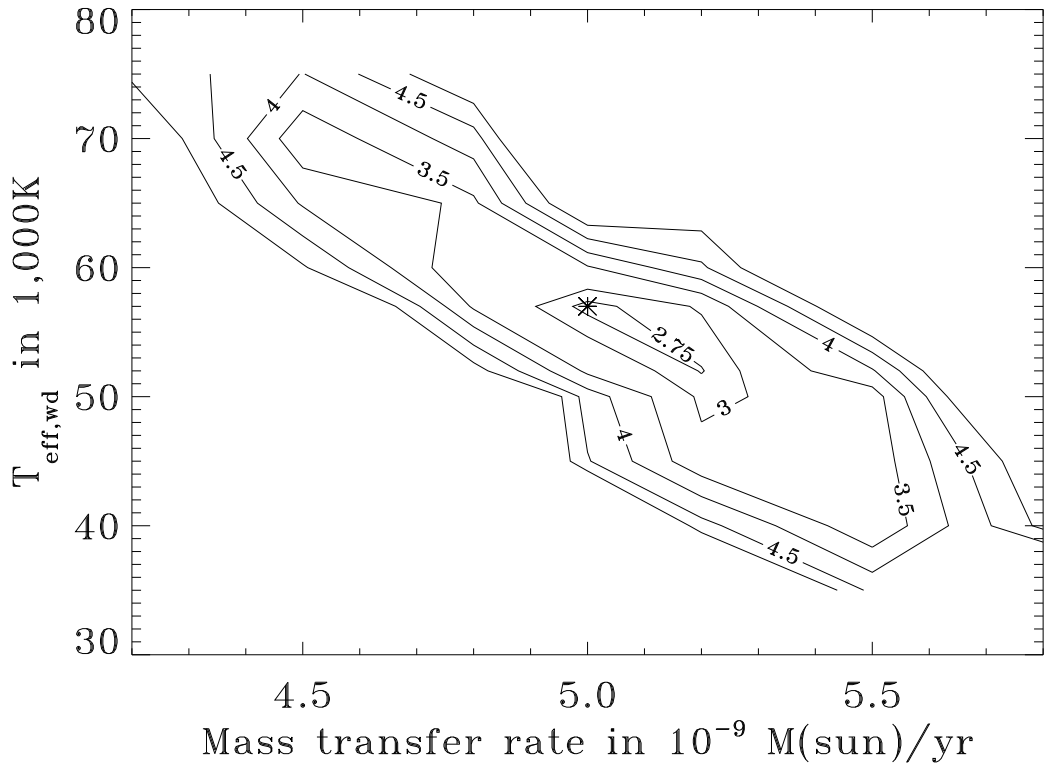}
\epsscale{1.00}
\vspace{5pt}
\figcaption{
Contour plot showing values of $\widetilde{\chi}^2$ in the $\dot{M}$, $T_{\rm eff,wd}$ plane
for the system models tabulated in Table~7. 
An asterisk marks the $\widetilde{\chi}^2$ value for the preferred solution.
See the text for a discussion.
\label{fg5}}
\end{figure}
%%%%%%%%%%%%%%%%%%%%%%%%%%%%%%%%%%%%%%%%%%%%%%%%%%%%%%%%%%%%%%%%%%%

\clearpage

%%%%%%%%%%%%%%%%%%%%%%%%%%%%%%%%%%%%%%%%%%%%%%%%%%%%%%%%%%%%%%%%%%%
\begin{figure}[tb]
\epsscale{0.97}
\plotone{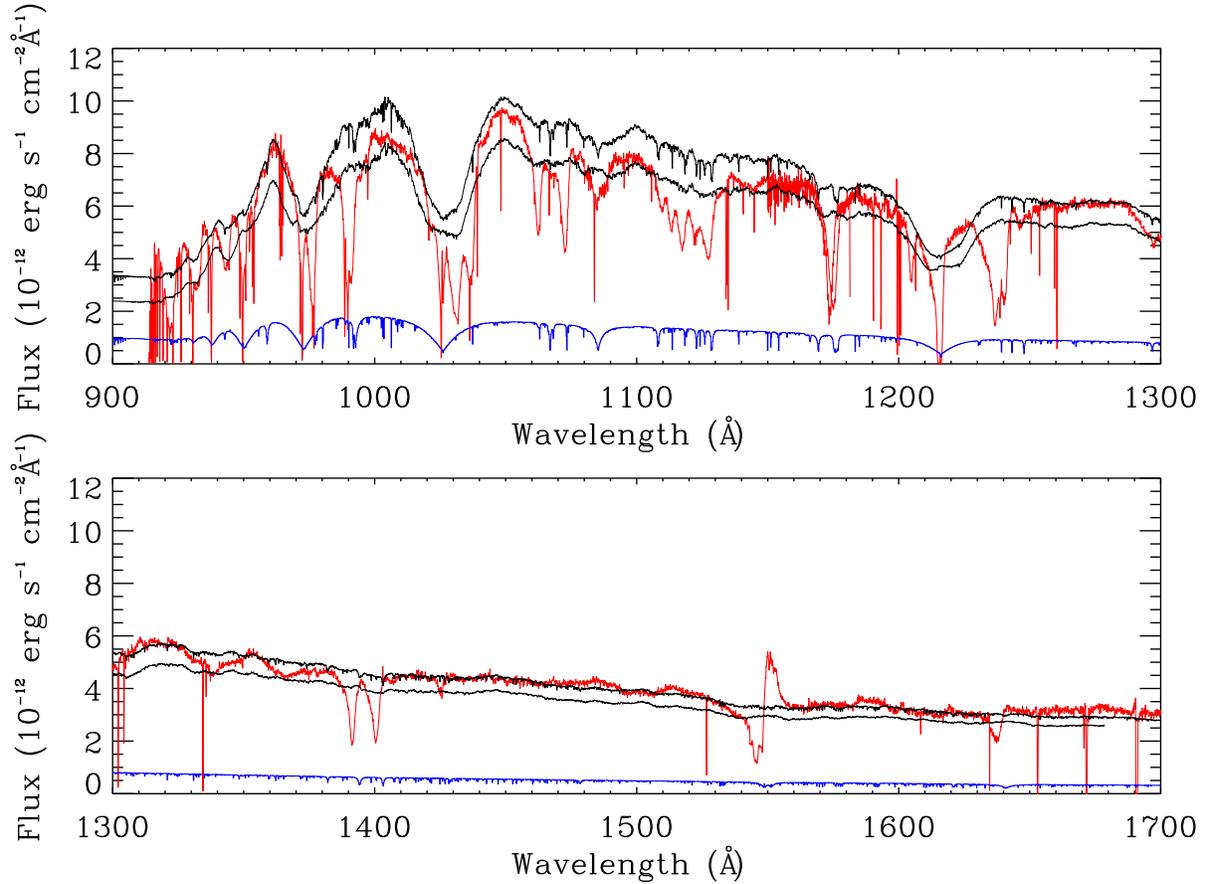}
\epsscale{1.00}
\vspace{5pt}
\figcaption{
As in Figure~3 but for the full wavelength range of the combined spectra.
The synthetic spectrum of the accretion disk alone is the next-to-top synthetic spectrum.
The uppermost system synthetic spectrum is the sum of the
accretion disk and the WD (blue line). The contributions of the secondary star and
accretion disk rim are negligible.
\label{fg6}}
\end{figure}
%%%%%%%%%%%%%%%%%%%%%%%%%%%%%%%%%%%%%%%%%%%%%%%%%%%%%%%%%%%%%%%%%%%

%%%%%%%%%%%%%%%%%%%%%%%%%%%%%%%%%%%%%%%%%%%%%%%%%%%%%%%%%%%%%%%%%%%
\begin{figure}[tb]
\epsscale{0.97}
\plotone{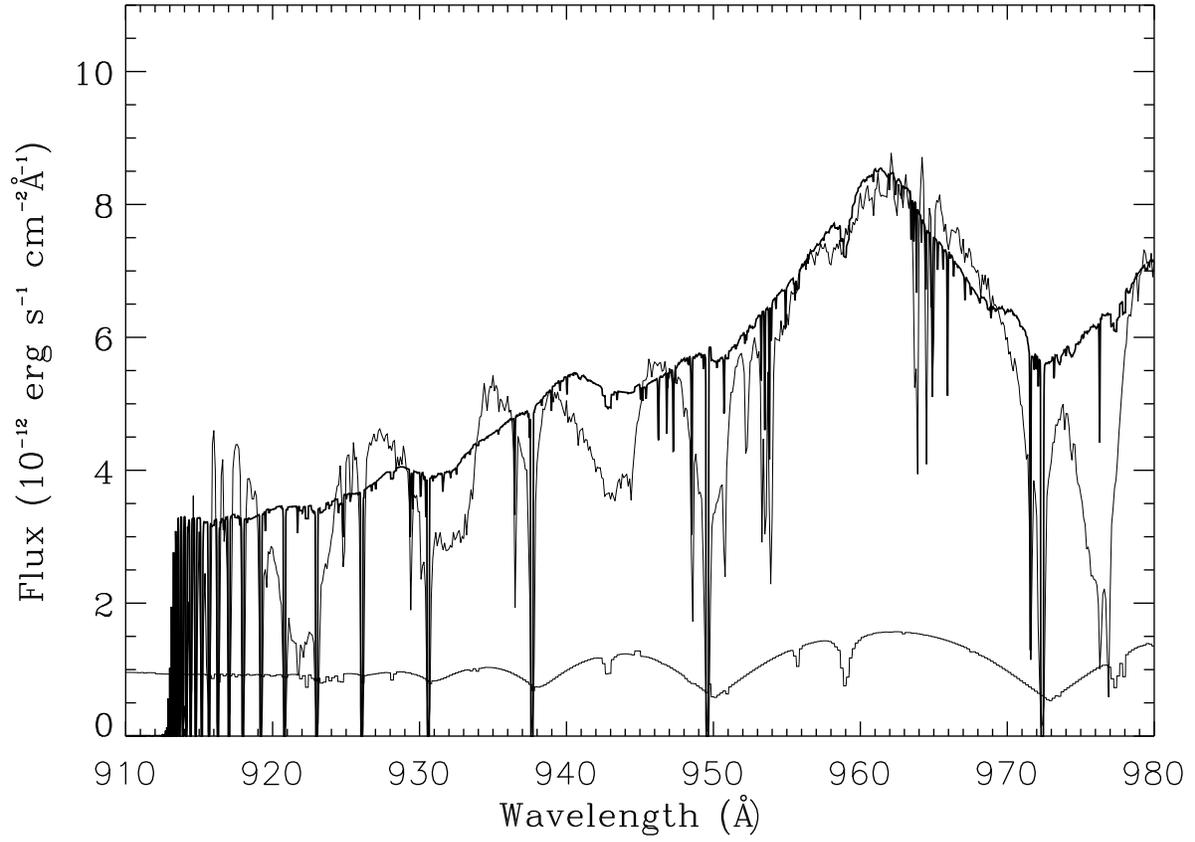}
\epsscale{1.00}
\vspace{5pt}
\figcaption{
The model of Figure~6 but with ISM correction. Note the cutoff at 912\AA.
\label{fg7}}
\end{figure}
%%%%%%%%%%%%%%%%%%%%%%%%%%%%%%%%%%%%%%%%%%%%%%%%%%%%%%%%%%%%%%%%%%%

\end{document}